\documentclass[aps,prl,floatfix,twocolumn,superscriptaddress]{revtex4-1}
\usepackage{amsmath}
\usepackage{amssymb}
\usepackage{amstext}
\usepackage{amsopn}
\usepackage{amsfonts}
\usepackage{amsxtra}
\usepackage[english]{babel}
\usepackage{graphicx}
\usepackage{float}
\usepackage{bm}
\usepackage{multirow}
\usepackage{dcolumn}
\usepackage{color}
\usepackage{hyperref}

\newcommand{\icm}{\ensuremath{~\textrm{cm}^{-1}}}

\begin{document}

\title{Infrared study of the multiband low-energy excitations of the topological antiferromagnet MnBi$_2$Te$_4$}

\author{Bing Xu} 
\affiliation{University of Fribourg, Department of Physics and Fribourg Center for Nanomaterials, Chemin du Mus\'{e}e 3, CH-1700 Fribourg, Switzerland}

\author{Y. Zhang} 
\affiliation{Sate Key Laboratory of Optoelectronic Materials and Technologies, School of Physics, Sun Yat-Sen University, Guangzhou, Guangdong 510275, China}

\author{E. H. Alizade}
\affiliation{Institute of Physics, Azerbaijan National Academy of Sciences, Baku AZ1143, Azerbaijan}

\author{Z. A. Jahangirli} 
\affiliation{Institute of Physics, Azerbaijan National Academy of Sciences, Baku AZ1143, Azerbaijan}
\affiliation{Baku State University, Z.Khalilov str. 23, AZ1148, Baku, Azerbaijan}

\author{F. Lyzwa} 
\author{E. Sheveleva} 
\author{P. Marsik} 
\affiliation{University of Fribourg, Department of Physics and Fribourg Center for Nanomaterials, Chemin du Mus\'{e}e 3, CH-1700 Fribourg, Switzerland}

\author{Y. K. Li} 
\author{Y. G. Yao} 
\author{Z. W. Wang} 
\email[]{zhiweiwang@bit.edu.cn}
\affiliation{Key Laboratory of Advanced Optoelectronic Quantum Architecture and Measurement, Ministry of Education, School of Physics, Beijing Institute of Technology, Beijing, 100081, China.}
\affiliation{Beijing Key Lab of Nanophotonics and Ultrafine Optoelectronic Systems, Beijing Institute of Technology, Beijing, 100081, China.}

\author{B. Shen} 
\email[]{shenbing@mail.sysu.edu.cn}
\affiliation{Sate Key Laboratory of Optoelectronic Materials and Technologies, School of Physics, Sun Yat-Sen University, Guangzhou, Guangdong 510275, China}

\author{Y. M. Dai} 
\affiliation{National Laboratory of Solid State Microstructures and Department of Physics, Nanjing University, Nanjing 210093, China}

\author{V. Kataev} 
\affiliation{Leibniz Institute for Solid State and Materials Research IFW Dresden, 01069 Dresden, Germany}

\author{M. M. Otrokov}
\affiliation{Centro de F\'{\i}sica de Materiales (CFM-MPC), Centro Mixto CSIC-UPV/EHU, 20018 Donostia-San Sebastián, Basque Country, Spain}
\affiliation{IKERBASQUE, Basque Foundation for Science, 48011 Bilbao, Basque Country, Spain}

\author{E. V. Chulkov}
\affiliation{Donostia International Physics Center, 20018 Donostia-San Sebastian, Basque Country, Spain}
\affiliation{Departamento de F\'{\i}sica de Materiales UPV/EHU, 20080 Donostia-San Sebastian, Basque Country, Spain}
\affiliation{Saint Petersburg State University, Laboratory of Electronic and Spin Structure of Nanosystems, 198504 Saint Petersburg, Russia}

\author{N. T. Mamedov} 
\email[]{n.mamedov@physics.ab.az}
\affiliation{Institute of Physics, Azerbaijan National Academy of Sciences, Baku AZ1143, Azerbaijan}

\author{Christian Bernhard} 
\email[]{christian.bernhard@unifr.ch}
\affiliation{University of Fribourg, Department of Physics and Fribourg Center for Nanomaterials, Chemin du Mus\'{e}e 3, CH-1700 Fribourg, Switzerland}

\date{\today}
%
%

\begin{abstract}
With infrared spectroscopy we studied the bulk electronic properties of the topological antiferromagnet MnBi$_2$Te$_4$ with $T_N \simeq 25~\mathrm{K}$. With the support of band structure calculations, we assign the intra- and interband excitations and determine the band gap of $E_g \approx$ 0.17~eV. We also obtain evidence for two types of conduction bands with light and very heavy carriers. The multiband free carrier response gives rise to an unusually strong increase of the combined plasma frequency, $\omega_{\mathrm{pl}}$, below 300~K. The band reconstruction below $T_N$, yields an additional increase of $\omega_{\mathrm{pl}}$ and a splitting of the transition between the two conduction bands by about 54 meV. Our study thus reveals a complex and strongly temperature dependent multi-band low-energy response that has important implications for the study of the surface states and device applications.
\end{abstract}


\maketitle

%
%
The focus in the research on topological quantum materials~\cite{Hasan2010RMP,Qi2011RMP,Haldane2017RMP,Tokura2019NRP} has recently moved on to systems with magnetic order which enable a variety of field-controlled quantum states~\cite{Wan2011PRB,Yu2010SC,Chang2013SC,Chang2015NM,Qi2008PRB,Essin2009PRL,Mong2010PRB,Xiao2018PRL,He2017SC}, like the quantum anomalous Hall (QAH) effect~\cite{Yu2010SC,Chang2013SC,Chang2015NM}, the topological axion state~\cite{Qi2008PRB,Essin2009PRL,Mong2010PRB,Xiao2018PRL}, and Majorana fermions~\cite{He2017SC,Qi2011RMP}. Such materials have been obtained, e.g. by creating heterostructures from magnetic and topological materials or by adding magnetic defects to topological materials. With the latter approach, the QAH effect has been realized for the first time in Cr-doped (Bi,Sb)$_2$Te$_3$ films~\cite{Chang2013SC}. The ideal candidates, however, are bulk topological materials with intrinsic magnetic order for which various problems inherent to thin film growth and defect engineering can be avoided.

A promising candidate is MnBi$_2$Te$_4$ (MBT) which is a topological insulator with A-type antiferromagnetic (AFM) order as predicted by theory ~\cite{Otrokov2019PRL,Li2019SA,Zhang2019PRL,Li2019PRB} and recently confirmed by experiements~\cite{Otrokov2019Nat,Zeugner2019CM,Vidal2019PRB,Lee2019PRR,Yan2019PRM,Cui2019PRB,Gong2019CPL,Yan2019PRB,Deng2020SC,Liu2020NM,Ge2020NSR,Hu2020NC,Wu2019SA,Chen2019NC}. Notably, the bulk AFM transition at $T_N \simeq$ 25~K has been predicted to strongly affect the electronic states at the (0001) surface, since it creates a gap on the Dirac cone~\cite{Otrokov2019PRL,Li2019SA,Zhang2019PRL,Otrokov2019Nat}. Moreover, for thin films the topological properties should depend on the number of MBT layers such that an axion insulator or a QAH insulator appears for even and odd numbers, respectively~\cite{Otrokov2019PRL,Li2019SA}. A quantized Hall conductance has indeed been observed in few layer MBT films~\cite{Deng2020SC,Liu2020NM,Ge2020NSR}, albeit only in magnetic fields of 5 -- 10 Tesla that change the magnetic order to a ferromagnetic one~\cite{Liu2020NM,Ge2020NSR}. The properties of the surface of MBT single crystals are also debated. For example, the formation of a gap below $T_N$ of the Dirac cone at the (0001) surface is seen in some angle-resolved photoemission spectroscopy (ARPES) studies ~\cite{Otrokov2019Nat,Zeugner2019CM,Vidal2019PRB,Lee2019PRR} but not in others~\cite{Nevola2020,Hao2019PRX,Chen2019PRX,Swatek2020PRB,Li2019PRX}. This calls for further studies of the surface structural and magnetic properties ~\cite{Hao2019PRX}. Likewise, the bulk-like low energy excitations and their modification in the AFM state remain to be fully understood.

Here we study the bulk electronic properties of MnBi$_2$Te$_4$ crystals with infrared (IR) spectroscopy. In combination with band-structure calculations, we assign the intra- and interband excitations and estimate the inverted bulk band gap and the chemical potential. We also study the excitations of the free carriers and determine their plasma frequency. The latter has a surprisingly low value and an unusual $T$ dependence, with a pronounced anomaly below $T_N$. We show that this anomalous behavior can be explained in terms of two conduction bands with largely different effective masses. Below $T_N$ we also identify a splitting of the transitions between the light and heavy conduction bands (by about 54~meV) that arises from the magnetic coupling between the conduction electrns and the localized Mn moments and agrees with the one seen with ARPES~\cite{Chen2019PRX,Swatek2020PRB,Li2019PRX,Estyunin2020}. This information about the multiband nature of the free carriers and their low-energy excitations is a prerequisite for understanding the plasmonic properties in the bulk as well as of the surface states and their eventual device applications. In the first place, it calls for attempts to reduce the defect concentration and thus the $n$-type doping such that a simpler single band picture applies.

\begin{figure}[tb]
\includegraphics[width=\columnwidth]{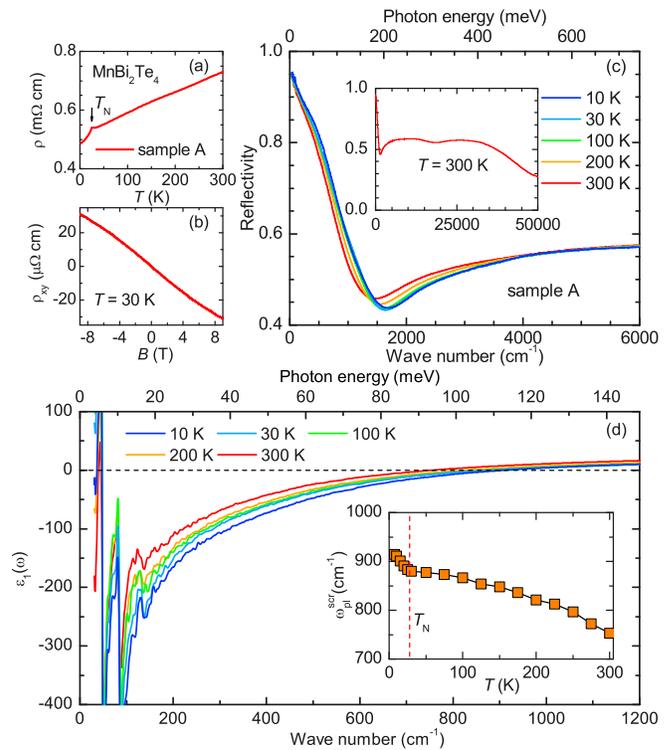}
\caption{ (color online) (a) $T$-dependent resistivity of the MnBi$_2$Te$_4$ sample A. The arrow marks the AFM transition at $T_N \simeq$ 25~K. (b) Hall resistance $R_{xy}$ of sample A at 30~K. (c) $T$ dependence of the reflectivity up to 6\,000\icm. Inset: Spectrum up to 50\,000\icm at 300~K. (d) $T$ dependence of the real part of the dielectric function $\varepsilon_1(\omega)$. Inset: Screened plasma frequency obtained from the zero crossing of $\varepsilon_1(\omega)$.}
\label{Fig1}
\end{figure}

%
%
Two batches of MBT single crystals were grown with a flux method~\cite{Chen2019NC} at Sun Yat-Sen University (Sample A) and Beijing Institute of Technology (Sample B). Both have a metallic in-plane resistivity with an anomaly around $T_N \simeq 25$~K, as shown in Fig.~\ref{Fig1}(a) for sample A. The negative Hall-resistivity $\rho_{xy}$ of sample A in Fig.~\ref{Fig1}(b) indicates electron-like carriers with a concentration of $n = 1.7 \times 10^{20}~\mathrm{cm}^{-3}$, in agreement with most previous studies~\cite{Lee2019PRR,Yan2019PRM,Cui2019PRB,Chen2019PRM,Li2020PCCP}. Details about the IR reflectivity measurements and the Kramers-Kronig analysis are given in section A of the supplemental material (SM).

%
%
Figure~\ref{Fig1}(c) shows for sample A the temperature ($T$) dependent reflectivity $R(\omega)$ up to 6\,000\icm. The inset shows the room temperature spectrum up to 50\,000\icm. Below about 1\,500\icm\ there is a sharp upturn of $R(\omega)$ toward unity that is characteristic of a plasma edge due to the itinerant carriers. This plasma edge shifts to higher frequency as the $T$ decreases, indicating an enhancement of free carrier density, $n$, or a reduction of effective mass, $m^{\ast}$. Very similar spectra have been obtained for sample B (see section B in the SM), the following discussion is therefore focused on sample A.

Fig.~\ref{Fig1}(d) displays the $T$ dependence of the real part of the dielectric function $\varepsilon_1(\omega)$. The spectra reveal an inductive behavior with a downturn of $\varepsilon_1(\omega)$ toward negative values at low frequency that is another hallmark of a metallic response. The sharp features at 47, 84 and 133\icm\ are IR-active phonons that are not further discussed here. The horizontal dashed line shows the zero crossing of $\varepsilon_1(\omega)$, which marks the screened plasma frequency $\omega_{\mathrm{pl}}^{\mathrm{scr}} = \omega_{\mathrm{pl}}/\sqrt{\varepsilon_{\infty}}$, where $\varepsilon_{\infty}$ is the high-frequency dielectric constant and $\omega_{\mathrm{pl}} = \sqrt{n e^2/\epsilon_0 m^\ast}$ is the free carrier plasma frequency. The inset details the $T$ dependence of $\omega_{\mathrm{pl}}^{\mathrm{scr}}$ which reveals an unusually large increase from about 750\icm\ at 300~K to 880\icm\ at 30~K. There is also a sudden, additional increase below $T_N$ to about 910\icm\ at 10~K which provides a first spectroscopic indication that the AFM order has a pronounced effect on the electronic properties.

\begin{figure}[tb]
\includegraphics[width=\columnwidth]{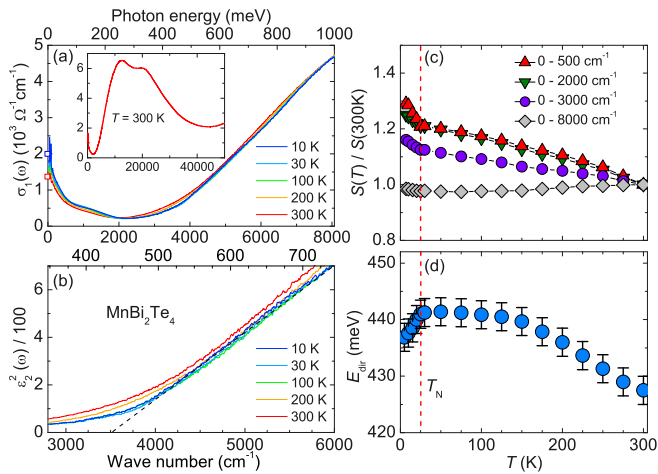}
\caption{ (color online) (a) $T$-dependent optical conductivity of MnBi$_2$Te$_4$ up to 8\,000\icm. The symbols on the $y$ axis denote $\sigma_{DC}$ at 10 and 300~K from the transport data in Fig.~\ref{Fig1}(a). Inset: Spectrum up to 50\,000\icm\ at 300~K. (b) $T$-dependent spectra of $\epsilon^2_2(\omega)$. The dashed line shows a linear extrapolation towards the zero crossing of $\epsilon^2_2(\omega)$ to obtain the onset of the direct interband transitions, $E_{dir}$. (c) $T$ dependence of the spectral weight for different cut-off frequencies. (d) $T$ dependence of $E_{dir}$.
}
\label{Fig2}
\end{figure}
Figure~\ref{Fig2}(a) displays the $T$ dependence of the real part of the optical conductivity $\sigma_1(\omega)$ up to 8\,000\icm. The inset shows the 300~K spectrum up to 50\,000\icm\ which is dominated by two interband transitions with bands around 12\,500 and 20\,000\icm, in agreement with Ref.~\cite{Zakir2019}. The optical response below 8\,000\icm\ consists of a Drude peak with a tail extending to about 2\,000\icm\ that is well separated from the onset of strong interband transitions above 3\,000\icm. The Drude peak grows upon cooling, consistent with the increase of $\omega^{\mathrm{scr}}_{\mathrm{pl}}$ in Fig.~\ref{Fig1}(d). The $dc$ conductivity data at 10 and 300~K from Fig.~\ref{Fig1}(a) (squares on the $y$-axis) agree with the zero-frequency extrapolation of $\sigma_1(\omega)$. The width of the Drude peak of about 500\icm\ is nearly $T$ independent and much larger than e.g. in Bi$_2$Te$_3$~\cite{Dordevic2013}. The scattering thus seems to be dominated by disorder effects e.g. due to Mn-Bi antisite defects~\cite{Zeugner2019CM}. The $T$ dependence of the onset of the strong interband transitions, $E_{dir}$, that are most likely direct transitions across the band gap, $E_g$, between the valence band (VB) and the conduction band (CB), has been obtained with a linear extrapolation of $\epsilon^2_2(\omega)$, as shown in Fig.~\ref{Fig2}(b). It increases toward low $T$, but decreases suddenly below $T_N$ [see Fig.~\ref{Fig2}(d)].

The spectral changes have been further analyzed by calculating the evolution of the spectral weight (SW), $S(\omega_c) = \int^{\omega_c}_0\sigma_{1}(\omega) d\omega$, for different cut-off frequencies $\omega_c$. Figure~\ref{Fig2}(c) shows the $T$ dependence of the ratio $S(\omega_c, T)/S(\omega_c, T = \mathrm{300~K})$ at representative cut-offs. At $\omega_c =$ 500 and 2\,000\icm, where the free carrier response dominates, the SW increases towards low $T$ and exhibits an additional upturn below $T_N$, in agreement with the trend of $\omega_{\mathrm{pl}}^{\mathrm{scr}}$ in Fig.~\ref{Fig1}(d). At the higher cutoffs, this increase becomes less pronounced until at $\omega_c =$ 8\,000\icm\ (1~eV) it is almost constant. This confirms that the SW redistribution is confined to energies below 1~eV.

\begin{figure*}[tb]
\includegraphics[width=2\columnwidth]{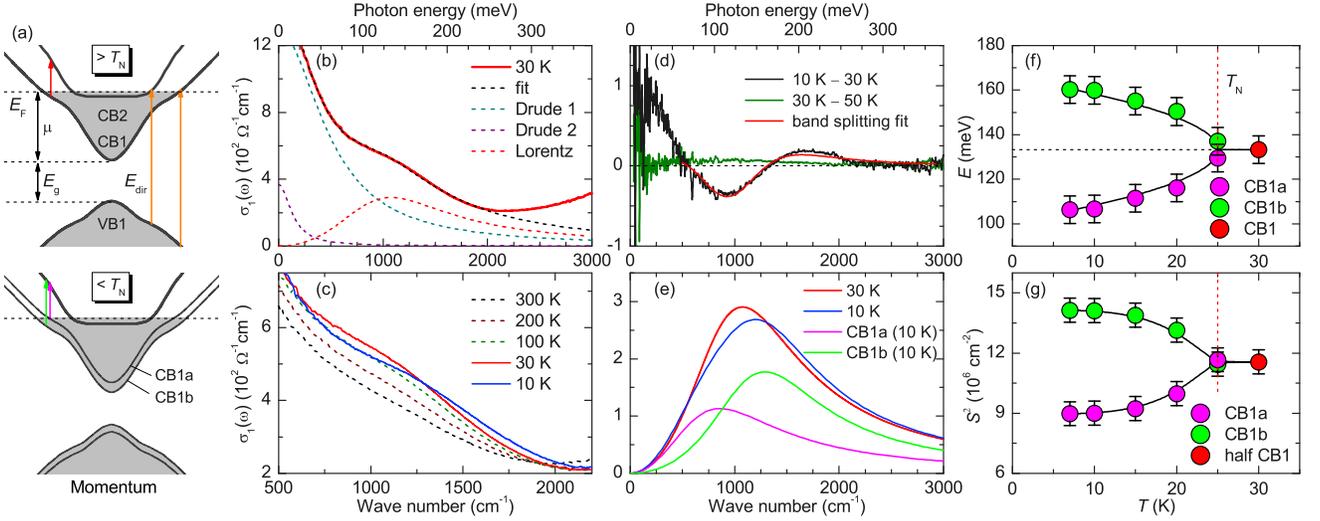}
\caption{ (color online) (a) Schematic of the band structure of MnBi$_2$Te$_4$ in the paramagnetic (upper) and AFM (lower panel) states. (b) Drude-Lorenz fit of the conductivity at 30~K around the lowest interband-transition. (c) $T$-dependent spectra showing the anomaly below $T_N \simeq$ 25~K. (d) Difference plots of $\sigma_1(\omega)$ and corresponding fit of the band splitting. (e) Lorenz fits of the low-energy interband transitions at 30 and 10~K  and of the split bands at 10~K. $T$ dependence of (f) the position and (g) the spectral weight of the split bands.
}
\label{Fig3}
\end{figure*}
Next, we analyze in more detail the response below 2\,000\icm\ which contains besides the Drude response a weak band due to a low-energy interband transition. This is evident in Fig.~\ref{Fig3}(b) which displays the $\sigma_1(\omega)$ spectrum at 30~K together with a Drude-Lorentz fit. It reveals a band centered around 1\,100\icm\ that overlaps with the tail of the Drude response. The fit function contains two Drude-terms with different plasma frequencies and scattering rates of $\omega_{\mathrm{pl},1} = 6215\icm$, $1/\tau_1 = 520\icm$ and $\omega_{\mathrm{pl},2} = 1870\icm$, $1/\tau_2 = 150\icm$, respectively. The band at 1\,100\icm\ is described by a Lorenz function. Details about the Drude-Lorentz analysis are given in section C of the SM.

Fig.~\ref{Fig3}(a) shows a schematics of the band structure in the vicinity of the chemical potential that is consistent with our optical data, with our band calculations along the $\Gamma$--Y direction (see setion E in the SM) and also with reported ARPES data~\cite{Chen2019PRX,Swatek2020PRB,Hao2019PRX}. In addition to a pair of conduction and valence bands that is forming an inverted band gap (CB1 and VB1), it contains a second conduction band (CB2) that is located slightly above CB1 and has a very flat bottom and thus a very large effective mass. As shown in the following, our optical data suggest that the chemical potential, $\mu$, is crossing both CB1 and CB2 (at low temperature). This assignment is consistent with the use of two Drude-peaks in fitting the low-energy response in the previous paragraph. It also accounts for the weak band around 1\,100\icm\ in terms of the interband transitions between CB1 and CB2 (red arrow). The optical excitations at higher energy involve transitions across the direct band gap $E_g$, from the VB to the empty states in CB1 and CB2, as illustrated in Fig.~\ref{Fig3}(a) by the orange arrows. Note that if $\mu$ would not be crossing CB2, the transition between the top of VB1 and the bottom of CB2, which are both rather flat and optically allowed, would give rise to a strong peak near $E_{dir}$ that is clearly not seen in the spectra of Fig.~\ref{Fig2}(a). On the other hand, a pronounced peak around 3\,350\icm\ (415~meV) has been observed in the corresponding spectra which were taken on the as grown surface of the same sample (see section D in the SM). This implies that for the as grown surface the chemical potential is somewhat lower, such that it falls below CB2. Such a reduction of the free carrier concentration might be caused, for example, by the localization of carries on extrinsic defects or by a lower concentration of intrinsic defects that are responsible for the $n$-type doping.

With this band assignment, we can estimate for the cleaved MBT surfaces the low-$T$ value of the chemical potential $\mu$ by using the expressions $\mu = \hbar^2k_F^2/2m^{\ast} = (\hbar^2/2m^\ast)(6\pi^2 n/g_sg_b)^{2/3}$, with the Fermi-vector $k_F$, the carrier density $n = \frac{1}{(2\pi)^3}\frac{4}{3}\pi k^3_F g_sg_b$, and the spin and band degeneracies $g_s = 2$ and $g_b = 2$~\cite{Zhang2019PRL,Li2019SA,Tang2016NP}. Using $n_{1} = 0.517 \times 10^{20}~\mathrm{cm}^{-3}$ and $n_{2} = 1.183 \times 10^{20}~\mathrm{cm}^{-3}$ (see section C in the SM), as well as $m^{\ast}_{1} = 0.12~m_e$ and $m^{\ast}_{2} = 3~m_e$ according to the band structure calculations (see section E in the SM), we derive $\mu_1 = 0.266$~eV and $\mu_2 = 0.019$~eV for CB1 and CB2, respectively. Accordingly, with an estimate of $E_{dir} \simeq$ 0.415~eV for the direct interband transition between VB1 and CB2 around the $\Gamma$ point, we derive a band gap of $E_g \approx E_{dir} + \mu_2- \mu_1 = 0.17 \pm 0.02$~eV at 30~K (as explained in section F of the SM, the largest uncertainty arises from the estimate of $\mu_2$), which agrees well with the reported values from band calculations and ARPES~\cite{Li2019SA,Zhang2019PRL,Li2019PRB,Otrokov2019Nat,Zeugner2019CM,Vidal2019PRB,Lee2019PRR,Hao2019PRX,Chen2019PRX,Swatek2020PRB}.

Next, we focus on the band reconstruction  below $T_N$, especially on the anomalous changes of the interband transition at 1\,100\icm\ which provide evidence for a magnetic splitting of CB1. In the paramagnetic state the $\sigma_1(\omega)$ spectra in Fig.~\ref{Fig3}(c) exhibit a monotonic increase in this frequency range that arises mainly from the growth of the Drude SW, as shown in Figs.~\ref{Fig1}(d) and \ref{Fig2}(c). Below $T_N$, this trend is suddenly interrupted, i.e. $\sigma_1(\omega)$ decreases from about 500 to 1\,200\icm\ whereas it gets anomalously enhanced between 1\,200 and 2\,000\icm. These anomalous changes, that are detailed in Fig.~\ref{Fig3}(d) in terms of the difference spectrum of $\sigma_1(\omega)$ at 30 and 10~K, are characteristic of a splitting of the conduction band CB1 into CB1a and CB1b, as indicated in the lower panel of Fig.~\ref{Fig3}(a). This band splitting, which is caused by the exchange interaction of the conduction electrons with the Mn moments which lifts the band degeneracy due to the unit cell doubling in the AFM state, is also seen in recent ARPES studies~\cite{Chen2019PRX,Swatek2020PRB,Estyunin2020,Li2019PRX}. Note that the magnetic splitting of CB2 is assumed to be much smaller and thus is neglected. This assumption is supported by ARPES data~\cite{Chen2019PRX,Swatek2020PRB,Estyunin2020,Li2019PRX}, and also by the comparison of the density of states at the Fermi-level derived from our optical data with the Korringa-slope of the ESR data in Ref.~\cite{Otrokov2019Nat}, as outlined in section H of the SM.
The red line in Fig.~\ref{Fig3}(d) confirms that the spectral changes below $T_N$ arise from a corresponding splitting of the interband transitions from CB1a to CB2 and CB1b to CB2. It has been obtained with the function: $\Delta\sigma_{1}(\omega)= L_a(\omega_a,\gamma_a,S_a) + L_b(\omega_b,\gamma_b,S_b) - L(\omega_0,\gamma,S)$, for which $L$ represents the Lorentz function and the subscripts $a$ and $b$ denote the interband transitions from the split bands, for which we assume $|\omega_a - \omega_0| = |\omega_b - \omega_0|$, $\gamma_a = \gamma_b = \gamma$ and $S^2_a + S^2_b = S^2$. The parameters in the paramagnetic state have been obtained from a Drude-Lorentz fit at 30~K. The contribution of the individual bands CB1a and CB1b are shown in Fig.~\ref{Fig3}(e). The red line in Fig.~\ref{Fig3}(d) shows that this band splitting model allows us to reproduce the $S$-shaped feature of $\sigma_1(\omega, \mathrm{10~K}) - \sigma_1(\omega, \mathrm{30~K})$. An additional contribution that arises from a much weaker and almost featureless $T$ dependent change of the background, that occurs also above $T_N$, has been corrected using the difference between 30 and 50~K (olive line). The full $T$ dependence of $\omega_a$ and $\omega_b$ below $T_N$ is displayed in Fig.~\ref{Fig3}(f). It shows that the band splitting amounts to 54~meV at 10~K and is almost symmetric with respect to the band position at 30~K. Fig.~\ref{Fig3}(g) displays the corresponding spectral weights $S^2_a$ and $S^2_b$ which exhibit a weak anisotropy that is consistent with the band splitting model. Note that the splitting of CB1 also accounts for the anomalous decrease of $E_{dir}$ below $T_N$, since it reduces $E_g$.

Finally, we return to the unusually large increase of $\omega_{\mathrm{pl}}^2$ towards low $T$ and its pronounced anomaly below $T_N$. The $\sim$ 20\% increase between 300 and 30~K can hardly arise from a volume contraction effect that would imply a giant expansion coefficient of $4 \times 10^{-3} \mathrm{K}^{-1}$. Likewise, the anomalous increase of $\omega_{\mathrm{pl}}^2$ below $T_N$ would require unrealistically large magneto-striction effects. Instead, we propose that the strong increase of $\omega_{\mathrm{pl}}^2$ toward low $T$ results from an exchange of conduction electrons between the light and very heavy states in CB1 and CB2~\cite{Drabble1958,Goldsmid1958,Chand1984PRB,Carter1970,Wieting1979}. Due to their largely different effective masses, the distribution of electrons is strongly dependent on the relative position of CB1 and CB2 with respect to the chemical potential. Accordingly, the $T$ dependence of the chemical potential accounts for the observed change of $\omega^2_{\mathrm{pl}}$ in the paramagnetic state (see section F in the SM). The anomalous increase of $\omega^2_{\mathrm{pl}}$ below $T_N$ requires in addition a small shift of the center of CB1a and CB1b against CB2 of $\sim$ 10~meV (see section G in the SM).
Note that the corresponding effect of the magnetic slitting of CB1a and CB1b is weaker and of the opposite sign  (see section G in the SM).

At last, we mention that for the majority of degenerate doped narrow gap semiconductors $\omega_{\mathrm{pl}}$ exhibits a much weaker $T$ dependence and usually decreases upon cooling due to the freeze-out of carriers. Interestingly, another rare exception, for which  $\omega_{\mathrm{pl}}^2$ exhibits a similarly strong increase toward low $T$, is Bi$_2$Te$_3$. While the samples studied in Ref.~\cite{Thomas1992PRB} where hole doped, in analogy to MBT, they may also have light and very heavy valence electrons.

%
%
In summary, we determined the bulk, optical properties of the AFM topological insulator MnBi$_2$Te$_4$. In combination with band structure calculations, we assigned the intra- and interband excitations and obtained a bulk band gap of $E_g \approx$ 0.17~eV. We also provided evidence for two conduction bands with largely different effective masses of 0.12 and 3~$m_e$ and chemical potentials of 0.266 and 0.019~eV (at 30~K). A $T$ dependent transfer of electrons between these conduction bands and the subsequent change of the average effective mass, can account for an unusually strong $T$ dependence of the free carrier plasma frequency, $\omega_{\mathrm{pl}}$. Below $T_N \simeq$ 25~K, we observed clear signs of a band reconstruction in terms of an additional, anomalous increase of $\omega_{\mathrm{pl}}$ and a splitting of the transition between the conduction bands. This detailed information about the bulk band structure and the multi-band charge carrier response is a prerequisite for the understanding of the plasmonic properties of the bulk and surface states and their device applications.

%
%
\begin{acknowledgments}
We acknowledge discussion with A. Akrap, G. Khalliulin and Z. Rukelj. The work in Fribourg was supported by the Schweizerische Nationalfonds (SNF) by Grant No. 200020-172611. V.K. acknowledges support by the Deutsche Forschungsgemeinschaft (DFG) through Grant No. KA1694/12-1. N.M. acknowledges the support of the Science Development Foundation under the President of the Republic of Azerbaijan (Grant No. E\.{I}F-BGM-4-RFTF-1/2017-21/04/1-M-02). The work at Beijing was supported by the Natural Science Foundation of China (NSFC Grant No. 11734003), the National Key R\&D Program of China (Grant No.2016YFA0300600), the Beijing Natural Science Foundation（Grant No. Z190006). Z.W. acknowledges the support from Beijing Institute of Technology Research Fund Program for Young Scholars. B.S. acknowledges the support of the Fundamental Research Funds for the Central Universities, Grant No. 19lgpy260. E.V.C. acknowledges Saint Petersburg State University (Grant No. ID 51126254).
\end{acknowledgments}

\clearpage
\appendix

\renewcommand{\thefigure}{S\arabic{figure}}
\setcounter{figure}{0}

\section{A: Reflectivity measurement, Kramers-Kronig analysis, and Hall resistivity data}
\begin{figure*}[tbh]
\includegraphics[width=0.85\textwidth]{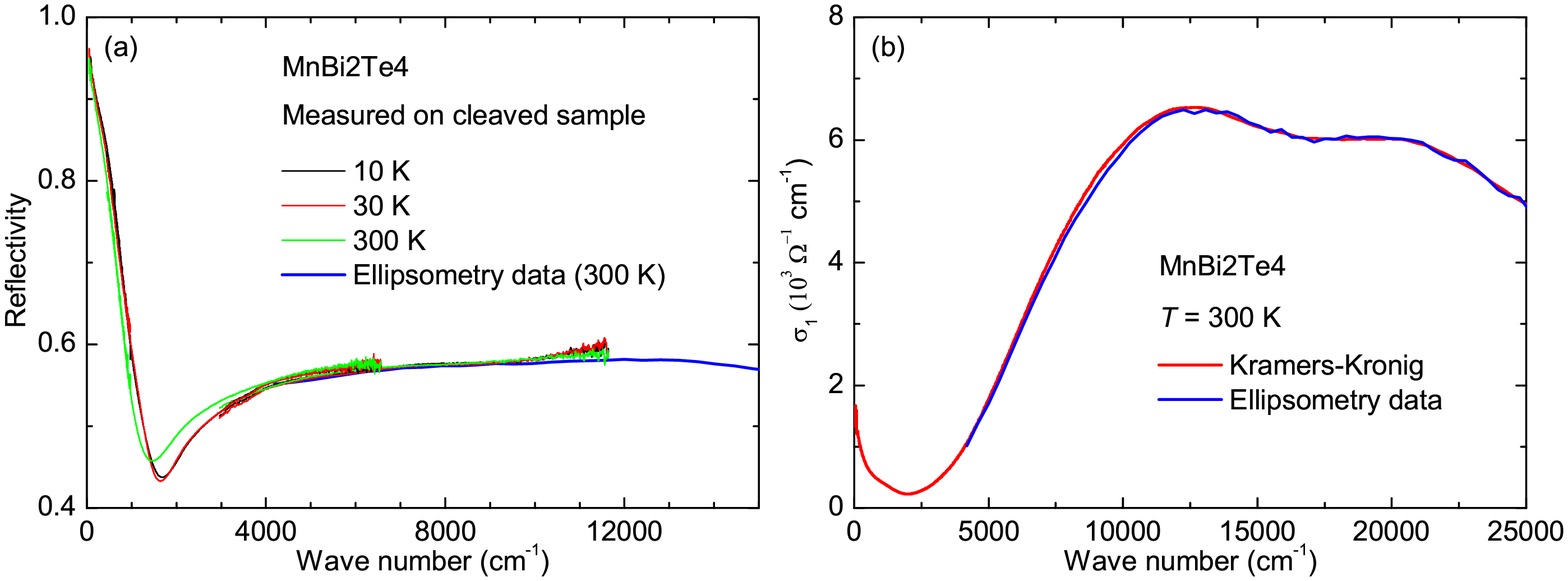}
\caption{ (color online) (a) Reflectivity spectra of MBT measured for different frequency ranges with a Fourier transform infrared reflectometer and with a grating-based spectroscopic ellipsometer (blue line). (b) Comparison of the $\sigma_1(\omega)$ spectrum at 300K in the range up to 25\,000\icm\ as obtained from the ellipsometry data (blue lines) and a Kramers-Kronlig analysis of the reflectivity data.}
\label{FigS1}
\end{figure*}
The in-plane reflectivity $R(\omega)$ of MnBi$_2$Te$_4$ was measured at a near-normal angle of incidence using a Bruker VERTEX 70v Fourier transform infrared spectrometer. An \emph{in situ} gold overfilling technique~\cite{Homes1993} was used to obtain the absolute reflectivity. As shown in Figure~\ref{FigS1}(a), the reflectivity spectra have been measured over a very broad frequency range from 30 to 12\,000\icm\ on a freshly cleaved sample surface by using a series of combinations of sources, beamsplitters and detectors. The reflectivity spectra at different temperatures from 300 to 7~K were collected with an ARS-Helitran cryostat. Figure~\ref{FigS1}(a) also shows the raw spectra measured for different frequency ranges at 10, 30 and 300~K. It is evident that all spectra exhibit the same temperature dependence and are overlapping very well in the region where the spectra have been connected to perform the Kramers-Kronig analysis. This confirms the accuracy and reproducibility of the measured reflectivity spectra.

The optical conductivity $\sigma_1(\omega)$ was obtained from a Kramers-Kronig analysis of $R(\omega)$~\cite{Dressel2002}. For the low-frequency extrapolation we used a Hagen-Rubens function $R = 1 - A\sqrt{\omega}$. On the high-frequency side, the ellipsometry data were used, for which the spectrum in the near-infrared to ultraviolet range (4\,000 -- 50\,000\icm) was measured at room temperature with a commercial ellipsometer (Woollam VASE), and the Kramers-Kronig analysis was anchored by the room temperature ellipsometry data, as shown by the spectra in Figure~\ref{FigS1}(b).

\begin{figure}[tbh]
\includegraphics[width=0.4\textwidth]{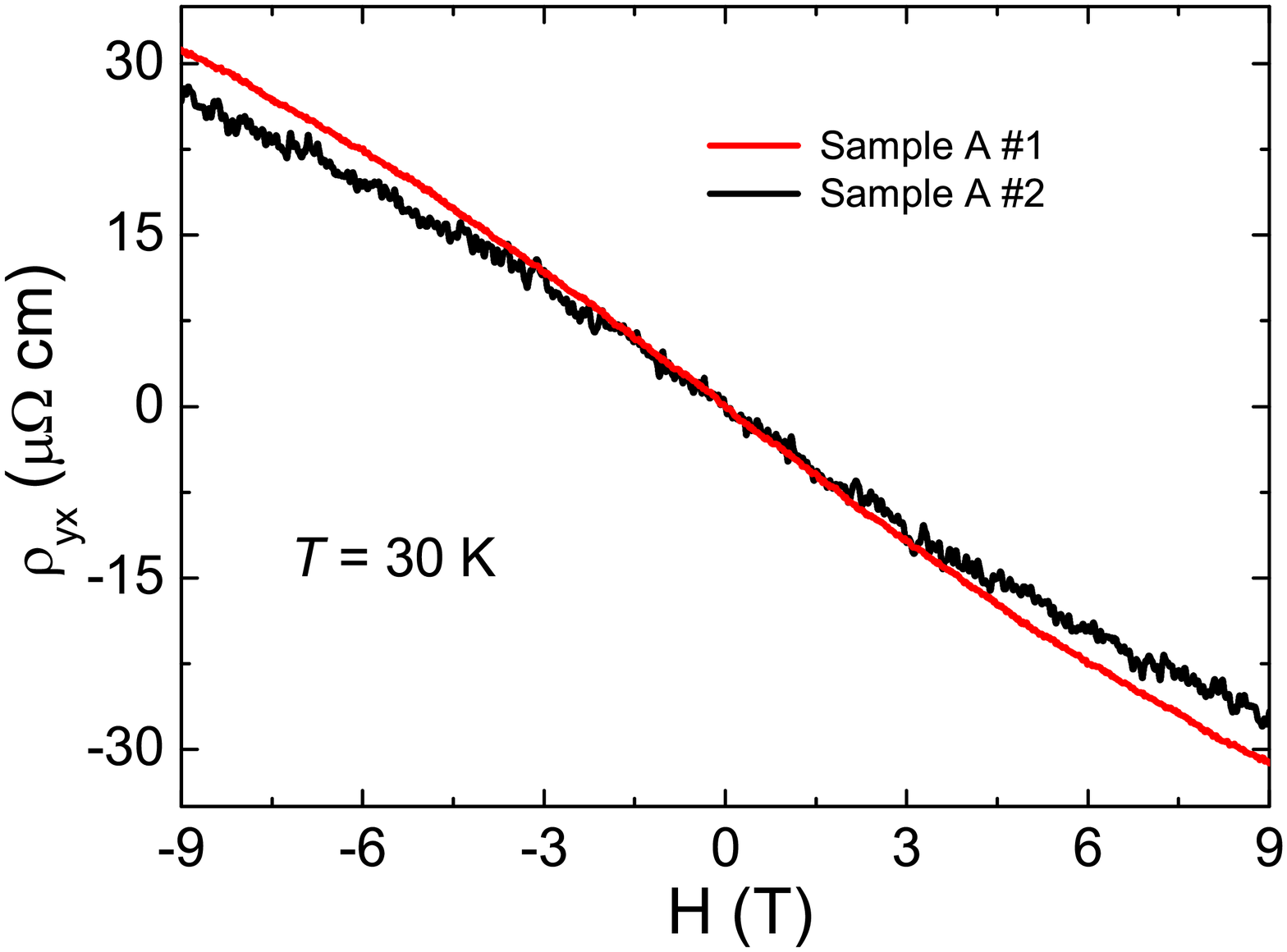}
\caption{ (color online) Hall resistivity data on different pieces of sample A for MnBi$_2$Te$_4$ at 30~K.}
\label{FigS2}
\end{figure}
In Fig.~\ref{FigS2}, we show the Hall resistivity data for MnBi$_2$Te$_4$ at 30~K that were taken on two different pieces of sample A from the same growth batch. Both samples exhibit a negative Hall resistivity with a linear slope from which we obtain conduction electron densities of $1.7 \times 10^{20}~\mathrm{cm}^{-3}$ and $1.9 \times 10^{20}~\mathrm{cm}^{-3}$, respectively. This confirms that the Hall-effect data are quite reproducible and accurate.

\section{B: Optical data of MnBi$_2$Te$_4$ collected on Sample B}

\begin{figure*}[tbh]
\includegraphics[width=\textwidth]{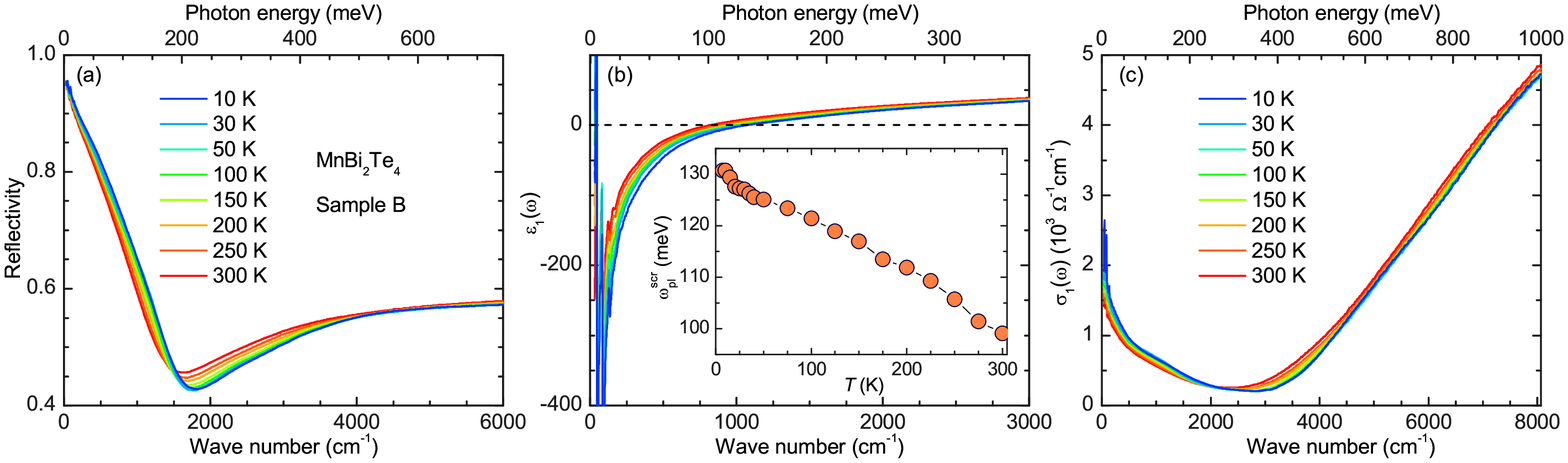}
\caption{ (color online) (a) Temperature dependence of the reflectivity spectra up to 6\,000\icm\ for MnBi$_2$Te$_4$ collected on sample B. (b) Temperature dependence of the real part of the dielectric function $\varepsilon_1(\omega)$. Inset: Screened plasma frequency of the free carriers obtained from the zero crossing of $\varepsilon_1(\omega)$. (c) Optical conductivity of MnBi$_2$Te$_4$ up to 8\,000\icm\ at different temperatures.}
\label{FigS3}
\end{figure*}
The optical data of MnBi$_2$Te$_4$ sample B were measured under the same conditions as reported for sample A in the main text. Fig.~\ref{FigS3} shows the temperature dependent optical data in the terms of the reflectivity [Fig.~\ref{FigS3}(a)], the real part of the dielectric function [Fig.~\ref{FigS3}(b)], and the real part of the conductivity [Fig.~\ref{FigS3}(c)]. The optical spectra are very similar to those of sample A as reported in the main text, further confirming the reproducibility and reliability of the measured spectra.

\section{C: Drude-Lorentz analysis}

We performed a quantitative analysis of the low-energy part of the $\sigma_1(\omega)$ spectra of sample A by fitting with the following Drude-Lorentz model.
\begin{equation}
\label{DrudeLorentz}
\sigma_{1}(\omega)=\frac{2\pi}{Z_{0}}\biggl[\sum_{j}\frac{\omega^{2}_{\mathrm{pl},j}}{\omega^{2}\tau_{j} + \frac{1}{\tau_{j}}} + \frac{\gamma\omega^{2}S^{2}}{(\omega^{2}_{0} - \omega^{2})^{2} + \gamma^{2}\omega^{2}}\biggr],
\end{equation}
where $Z_0$ is the vacuum impedance. The first term with a sum of two Drude terms describes the response of the itinerant carriers in the two conduction bands CB1 and CB2, each characterized by a plasma frequency $\omega_{\mathrm{pl},j}$ and a scattering rate $1/\tau_j$. The second term denotes a Lorentz oscillator with the resonance frequency $\omega_{0}$, line width $\gamma$ and oscillator strength $S$, that accounts for the interband transitions between CB1 and CB2.

\begin{figure*}[tbh]
\includegraphics[width=0.9\textwidth]{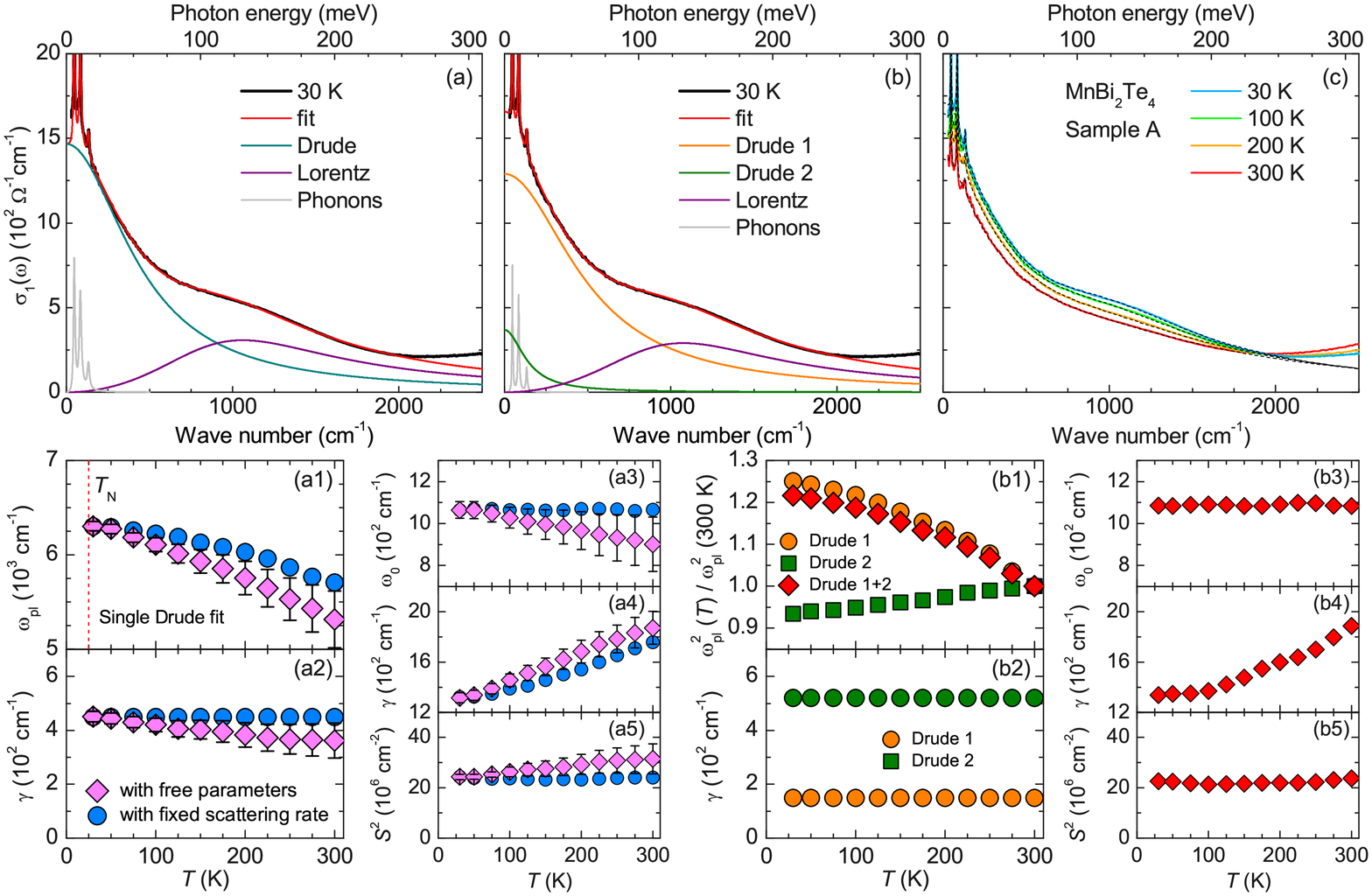}
\caption{ (color online) (a) Fit of the conductivity at 30~K with the function in equation (1) using only a single Drude term. (b) Corresponding fit of the conductivity at 30~K with two Drude terms. (c) Fits with two Drude terms (dashed lines) to the $\sigma_1(\omega)$ spectra (colored lines) at representative temperatures above $T_N$. (a1--a5) $T$ dependence of the parameters of the Drude- and Lorentz terms as obtained with a single Drude term for which the scattering rate is either free (diamonds) or fixed (circles) at the value obtained at 30K. (b1--b5) Corresponding $T$ dependence of the fit parameters obtained with two Drude terms.}
\label{FigS4}
\end{figure*}

In a first attempt, trying to reduce the fitting parameters, we used a single Drude band and the Lorentz band. Figures~\ref{FigS4}(a) shows the corresponding Drude-Lorentz fit (red line) to the low-energy part of the $\sigma_1(\omega)$ spectrum at at 30~K (black curve). Also shown are the contributions of the Drude term (cyan line) and the Lorentz term (purple line) due to the free carriers and the low energy interband transition. An additional contribution from infrared active phonons is shown by the gray line. The obtained plasma frequency and scattering rate at 30~K are $\omega_{\mathrm{pl}} = 6300\icm$ and $1/\tau = 450\icm$, respectively. Using the squared plasma frequencies, $\omega^2_{\mathrm{pl}} = n e^2/\epsilon_0 m^\ast$, as a measure of the ratio of the free carrier density $n$, and the effective mass $m^\ast$ and with $n = 1.7 \times 10^{20}~\mathrm{cm}^{-3}$, as obtained from the Hall data at 30~K, we derive an effective mass of the conduction electrons of $m^{\ast} \simeq 0.38~m_e$. Such a large value of the effective mass disagrees with the band structure calculations in section E which predict a much smaller effective mass of the lowest conduction band CB1 of $m_1^\ast = 0.12~m_e$. As discussed in the main text, this discrepancy can be resolved by taking into account an additional contribution from the second conduction band CB2 that has very heavy carriers with $m_2^\ast = 3~m_e$.

To account for the presence of the two conduction bands CB1 and CB2, we have fitted the spectra with the function in equation (\ref{DrudeLorentz}) using two Drude-terms. Figure~\ref{FigS4}(b) shows that the two-Drude-model fit (red line) to the spectrum at 30~K (black line) yields a better description in very low energy range than the single-Drude fit in Figure~\ref{FigS4}(a). As is also described in the main text, the obtained parameters of the two Drude terms are $\omega_{\mathrm{pl},1} = 6215\icm$, $1/\tau_1 = 520\icm$ and $\omega_{\mathrm{pl},2} = 1870\icm$, $1/\tau_2 = 150\icm$, for CB1 and CB2, respectively. With $n_{1} + n_{2} = n = 1.7 \times 10^{20}~\mathrm{cm}^{-3}$ taken from the Hall-data and the combined plasma frequency of CB1 and CB2, $\omega_{\mathrm{pl}} = \sqrt{\omega^2_{\mathrm{pl},1} + \omega^2_{\mathrm{pl},2}} = \sqrt{n_{1}e^2/\epsilon_0 m^{\ast}_{1} + n_{2}e^2/\epsilon_0 m^{\ast}_{2}} = 6490\icm$, where $m^{\ast}_{1} = 0.12~m_e$ and $m^{\ast}_{2} = 3~m_e$ according to the band structure calculations in section E, we derive the carrier concentrations of $n_{1} \simeq 0.517 \times 10^{20}~\mathrm{cm}^{-3}$ and $n_{2} \simeq 1.183 \times 10^{20}~\mathrm{cm}^{-3}$ for CB1 and CB2, respectively. Alternatively, with the partial plasma frequency $\omega_{\mathrm{pl},1} = \sqrt{ n_{1}e^2/\epsilon_0 m^{\ast}_{1}} = 6215\icm$ and $\omega_{\mathrm{pl},2} = \sqrt{n_{2}e^2/\epsilon_0 m^{\ast}_{2}} = 1870\icm$, we obtain $n_{1} \simeq 0.518 \times 10^{20}~\mathrm{cm}^{-3}$ and $n_{2} \simeq 1.172 \times 10^{20}~\mathrm{cm}^{-3}$ for CB1 and CB2, respectively, and thus $n_{1} + n_{2} = 1.69 \times 10^{20}~\mathrm{cm}^{-3} \approx n = 1.7 \times 10^{20}~\mathrm{cm}^{-3}$. The good agreement of both estimates confirms the self-consistency of our fitting approach with two Drude bands from two conduction bands with light and very heavy electrons, the effective mass from band calculations, and the carrier concentration from the Hall data.

Next, we discuss the $T$ dependence of the obtained fit parameters in the paramagnetic state above $T_N$ for the single-Drude as well as the two-Drude models.  Figs.~\ref{FigS4}(a1--a5) show the $T$ dependence of the fit parameters for the single Drude model. It yields a nearly $T$ independent scattering rate that is also evident from the nearly equal width of the $\sigma_1(\omega)$ spectra in Fig.~\ref{FigS4}(c). Accordingly, to reduce the number of fit parameters and to avoid a mixing of spectral weight as the width of the latter increases toward high $T$, we have fitted the $T$ dependent spectra with a fixed scattering rate as obtained at 30 K. Figures~\ref{FigS4}(b1--b5) show the corresponding $T$ dependent fit parameters as obtained with the two-Drude model where again the scattering rates have been fixed to the value at ~30 K. Figure~\ref{FigS4}(b1) shows the $T$ dependence of the normalized Drude weights $\omega^2_{\mathrm{pl}}(T)/\omega^2_{\mathrm{pl}}(T = \mathrm{300~K})$. As described in the main text, the total Drude weight increases towards low $T$ and grows by about 20\% between 300 and ~30 K. Notably, the partial Drude weights of CB1 and CB2 exhibit opposite  $T$ dependent trends, i.e. whereas the Drude weight of CB1 increases toward $T_N$ the one of CB2 decreases. This behavior is consistent with our interpretation, as described in the main text and in section F, that electrons are transferred from CB2 to CB1 as $T$ is reduced.

\section{D: Additional peak around 3\,350\icm in the spectra on the as-grown surface}

\begin{figure*}[tbh]
\includegraphics[width=\textwidth]{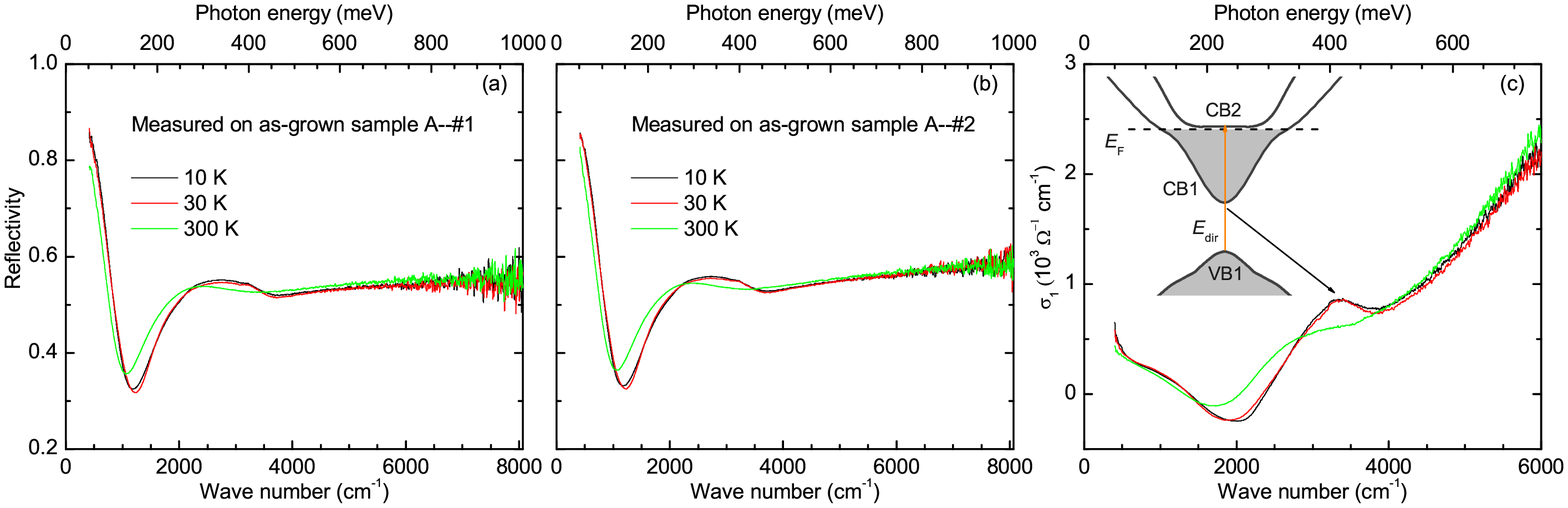}
\caption{ (color online) (a--b) Reflectivity spectra in the mid-infrared frequency range of MnBi$_2$Te$_4$ measured on as-grown surfaces of sample A. (c) Corresponding spectra of the optical conductivity of MnBi$_2$Te$_4$ as derived from a preliminary Kramers-Kronig analysis using reflectivity spectra that cover only a limited frequency range. Inset: Schematic of the band structure of MnBi$_2$Te$_4$ around the Fermi-level showing the origin of the pronounced peak around 3\,350\icm.}
\label{FigS5}
\end{figure*}
Figures~\ref{FigS5}(a) and ~\ref{FigS5}(b) show the raw spectra measured in the mid-infrared range  on an as-grown surface for two different pieces of sample A. In both cases we observe a strong additional peak around 3\,350\icm\ that does not show up in the corresponding spectra taken on cleaved surfaces of the same sample A. This peak has been reproduced on several as grown surfaces and apparently is intrinsic to the sample properties rather than just an artefact or a dirt effect. Figure~\ref{FigS5}(c) shows the corresponding optical conductivity  $\sigma_1(\omega)$ at the as grown surface that has been obtained from a preliminary Kramers-Kronig analysis using the reflectivity data that cover only a limited frequency range. It reveals a pronounced peak peak around 3\,350\icm. This additional peak can be naturally explained, as shown by the sketch in the inset of Fig.~\ref{FigS5}(c), if the Fermi level at the as-grown surface is slightly reduced as compared to the one of the cleaved surface, such that it is located below CB2. In this case, the interband transitions between the top of VB1 and the bottom of CB2 are allowed and yield a pronounced peak since both bands are rather flat giving rise to a high joint density of states. In the future, this assignment could be confirmed with optical studies on cleaved surface of samples with a much lower carrier concentration than for our samples A and B.

\section{E: Effective mass from band structure calculations and band degeneracy}

\begin{figure}[tbh]
\includegraphics[width=0.5\textwidth]{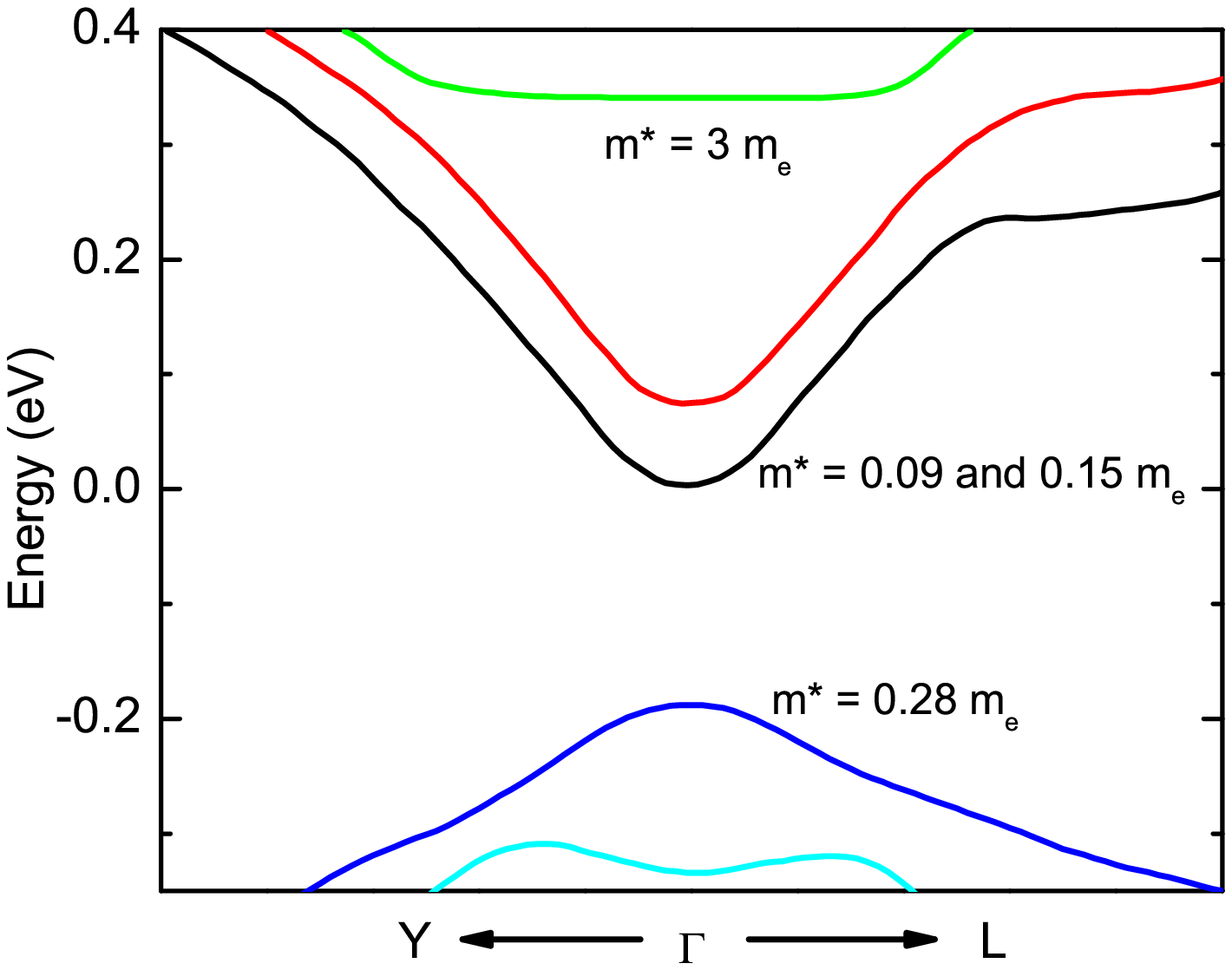}
\caption{ (color online) Schematics of the calculated bulk band dispersion of MnBi$_2$Te$_4$ along $\Gamma$Y and $\Gamma$L in the AFM state.}
\label{FigS6}
\end{figure}
As shown in Fig.~\ref{FigS6}, the calculations of the band structure (details are described in Ref.~\cite{Zakir2019}) in the AFM state yield two ``split-type'' light electron conduction bands with electron effective masses of 0.09 and 0.15~$m_e$ for the lower (black) and upper one (red), respectively. In the paramagnetic state these two bands are degenerate and therefore described in the manuscript as a single band with an average effective mass of 0.12~$m_e$ (CB1).
Note that this double degeneracy ($g_b = 2$) is a consequence of the combined inversion and time-reversal symmetry ($P\Theta$ symmetry) as described in Ref.~\cite{Zhang2019PRL,Li2019SA,Tang2016NP} and applies to all bands in the paramagnetic state. The splitting of CB1 in the AFM state, into a lower band with  0.09~$m_e$ (CB1b) and an upper one with  0.15~$m_e$ (CB1a), which is predicted to amount to approximately 80~meV, is a consequence of the broken $P$ symmetry that arises from the exchange interaction of the conduction electrons with the localized Mn spins for which the unit cell in the A-type AFM state is doubled along the c-axis. This magnetic interaction thus lifts the band degeneracy ($g_b = 1$) but maintains the spin degeneracy ($g_s = 2$) of CB1a and CB1b. There is also a third band of heavy electrons with very small dispersion and a very large effective mass around 3~$m_e$ that is denoted in the main test as CB2. Since the exchange interaction of its carriers with the localized Mn moments is assumed to be very weak we describe it even in the AFM state in terms of a single band with $g_b = 2$ and $g_s = 2$. Finally, the estimate of the hole mass of the valence band is about 0.28~$m_e$.

Effective masses of electrons and holes were estimated by fitting parabola to the calculated dispersion of the electron bands around the $\Gamma$ point in the directions $\Gamma$--Y and $\Gamma$--L  of the Brillouin zone of MnBi$_2$Te$_4$. Note that the $\Gamma$--Y direction lies exactly in the layer plane, as is required to account for the optical transitions in the experimental data for which the polarization vector of the incident light is parallel to the layer plane (perpendicular to the $c$-axis) of the MBT crystal.

\section{F: Multi-band approach to describe the $T$ dependent redistribution of electrons from CB2 to CB1}

The density of states $D(E)$ of the conduction electrons of an $n$-type doped three dimensional material is given by
\begin{equation}
D(E)= \frac{g_sg_b}{4\pi^2} (\frac{2m^{\ast}}{\hbar^2})^{3/2} \sqrt{E -E_c},
\end{equation}
where $E_c$ is the conduction band edge. At finite temperature the Fermi-Dirac distribution is given by
\begin{equation}
f(E)= \frac{1}{1 + e^{(E-E_F)/kT}},
\end{equation}
where $E_F$ = $\mu$ ($T =$ 0~K) is the Fermi energy and $\mu$ the chemical potential. Accordingly, the carrier density at finite temperature is given by
\begin{equation}
n = \frac{g_sg_b}{4\pi^2} (\frac{2m^{\ast}}{\hbar^2})^{3/2} \int^{\infty}_{E_c}\frac{\sqrt{E -E_c}}{1 + e^{(E-E_F)/kT}}dE.
\end{equation}
With the definitions $\nu = (E_F - E_c)/kT \equiv \mu/kT$ and $y = (E -E_c)/kT$, this integral can be rewritten in the normalized form:
\begin{equation}
\label{Fermi-Dirac-int}
n = n_c\frac{g_sg_b}{\sqrt{\pi}} \int^{\infty}_{0}\frac{\sqrt{y}}{1 + e^{y-\nu}}dy \equiv n_c\frac{g_sg_b}{\sqrt{\pi}} F_{1/2}(\nu).
\end{equation}
where the ``effective'' density of states in the conduction band is
\begin{equation}
n_c = 2(\frac{m^{\ast}kT}{2\pi\hbar^2})^{3/2} = 2.51 \times 10^{19} \times [\frac{m^{\ast}}{m_e}]^{3/2} \times [\frac{T}{300~\mathrm{K}}]^{3/2} \mathrm{cm}^{-3},
\end{equation}
and $F_{1/2}(\nu)$ is known as the Fermi-Dirac integral of order 1/2 (referring to the $y^{1/2}$ in the numerator).

\begin{figure}[tbh]
\includegraphics[width=0.5\textwidth]{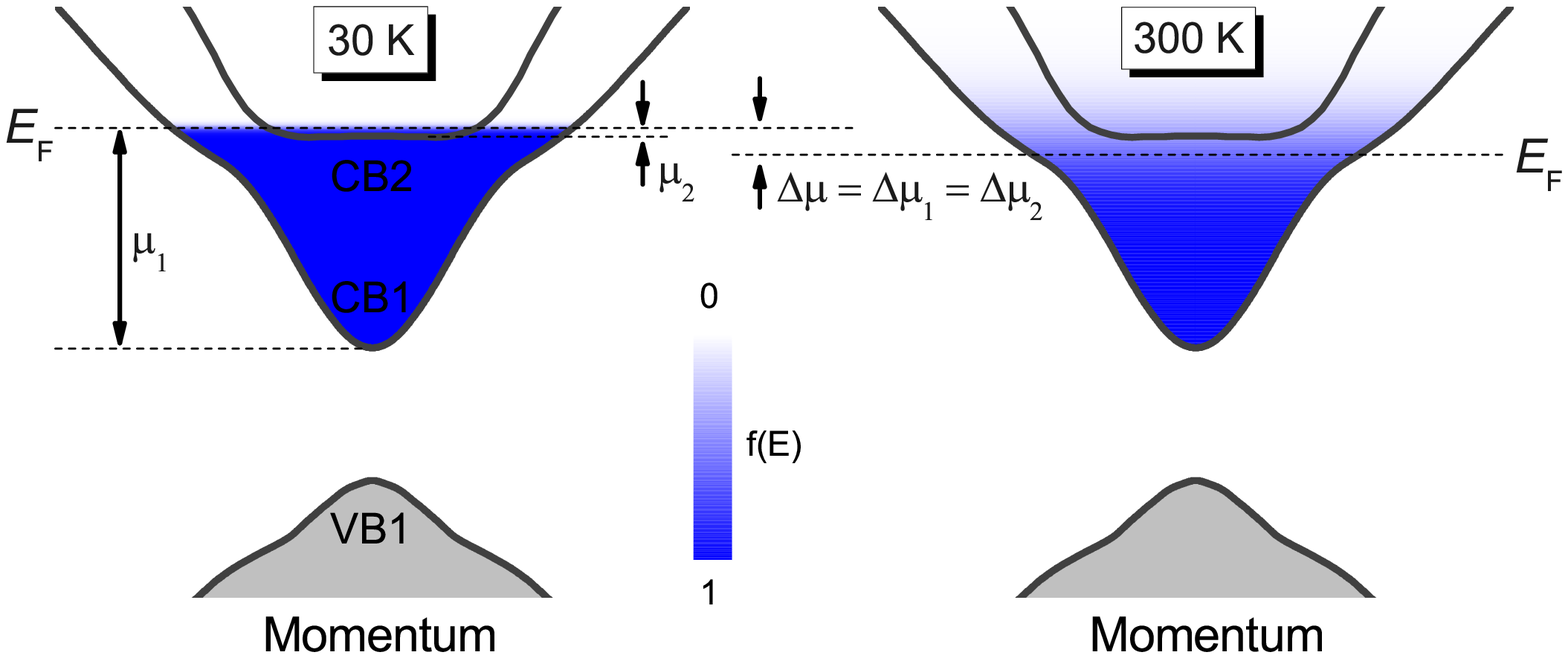}
\caption{ (color online) Schematics of the thermal broadening of the Fermi-edge and the related $T$ dependent shift of the chemical potential between 30 and 300~K.}
\label{FigS7}
\end{figure}
\begin{figure*}[tbh]
\includegraphics[width=0.8\textwidth]{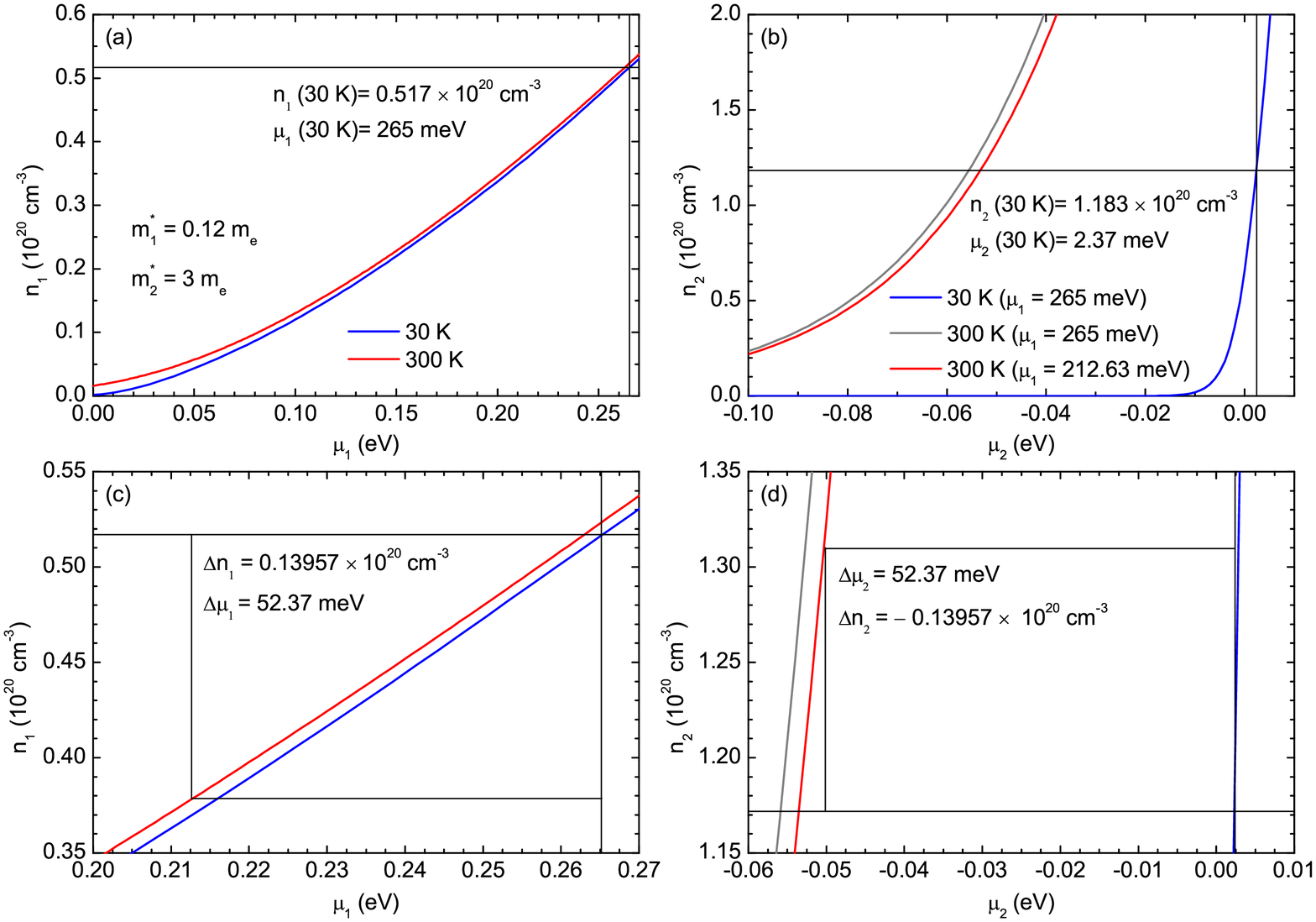}
\caption{ (color online) (a) Carrier density $n_1$ of the band CB1 as a function of $\mu_1$. (b) Carrier density $n_2$ of the band CB2 as a function of $\mu_2$. Enlarged view of (a) and (b) showing the $T$ dependent redistribution of electrons between CB2 and CB1.}
\label{FigS8}
\end{figure*}
\begin{figure*}[tbh]
\includegraphics[width=\textwidth]{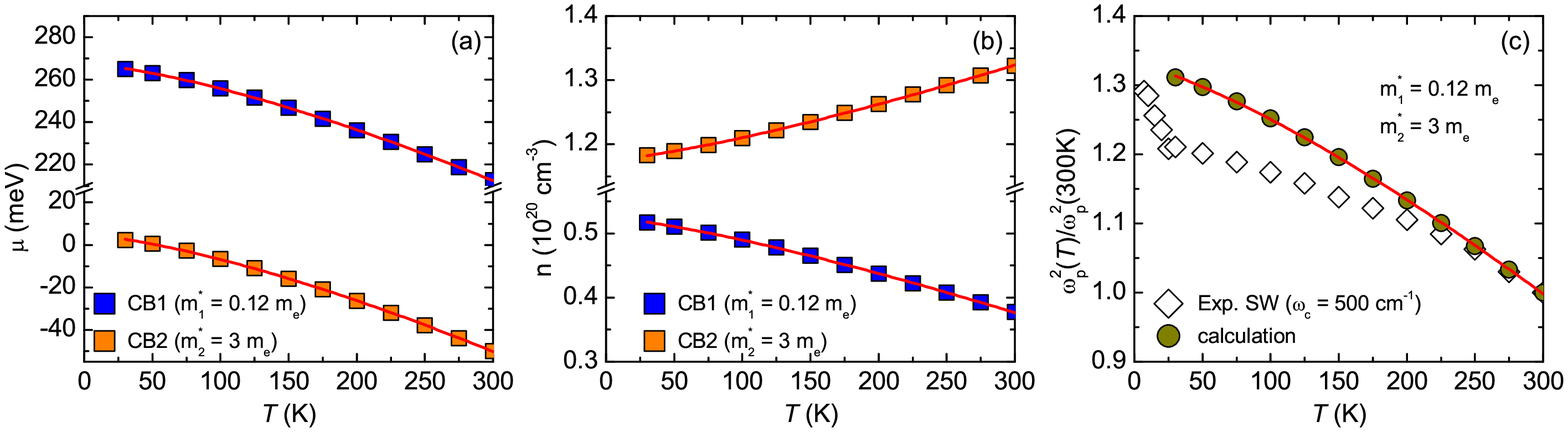}
\caption{ (color online) (a) $T$ dependent shift of the chemical potential for CB1 and CB2. (b) $T$ dependence of the carrier densities of CB1 and CB2. (c) $T$ dependence of the ratio $\omega^2_{\mathrm{pl}}(T)/\omega^2_{\mathrm{pl}}(\mathrm{300~K})$.}
\label{FigS9}
\end{figure*}
\begin{figure*}[tbh]
\includegraphics[width=0.8\textwidth]{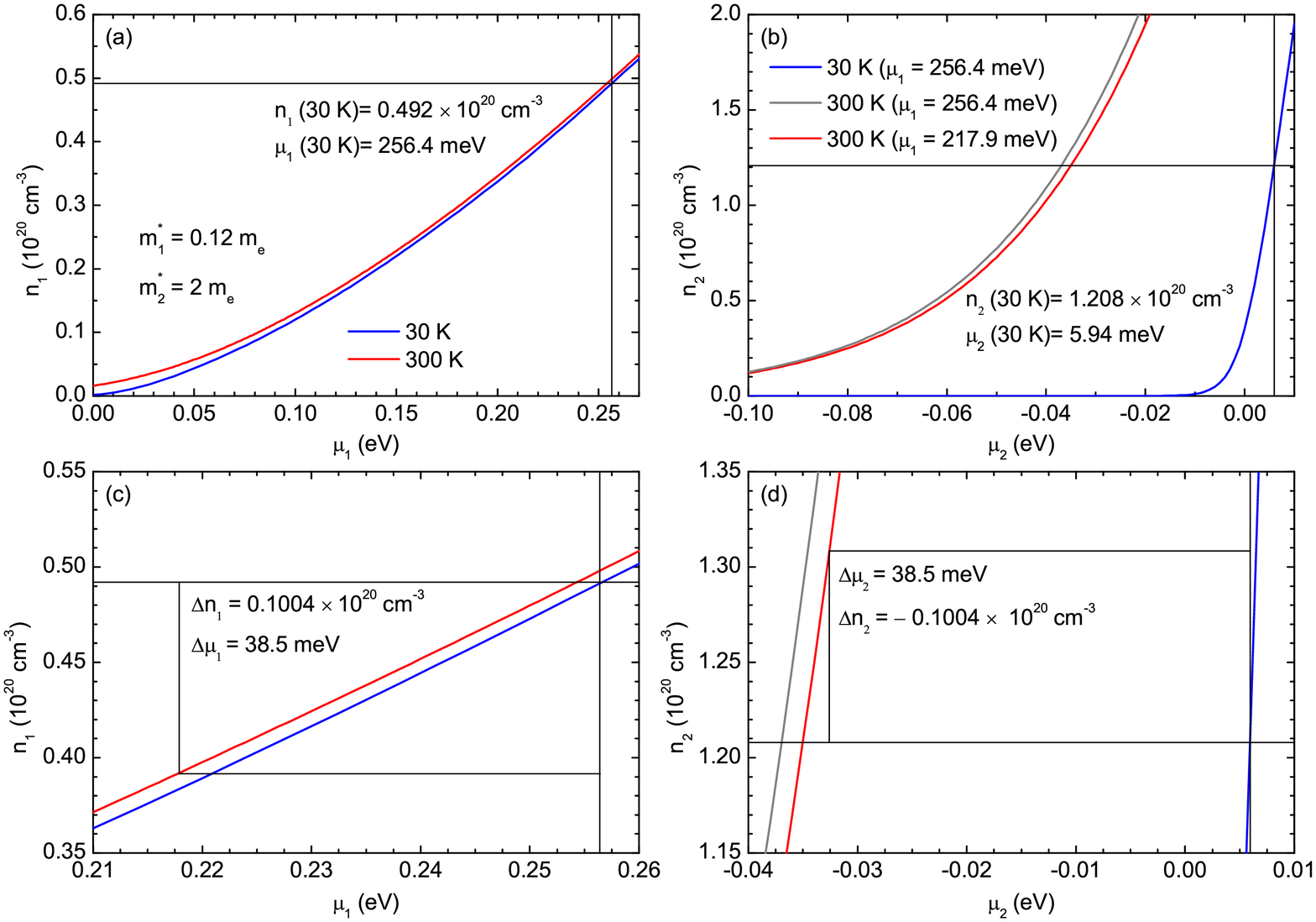}
\caption{ (color online) (a) Carrier density $n_1$ of CB1 as a function of $\mu_1$. (b) Carrier density $n_2$ of CBa as a function of $\mu_2$. Enlarged view of (a) and (b) showing the $T$ dependent redistribution of electrons between CB2 and CB1.}
\label{FigS10}
\end{figure*}
\begin{figure*}[tbh]
\includegraphics[width=\textwidth]{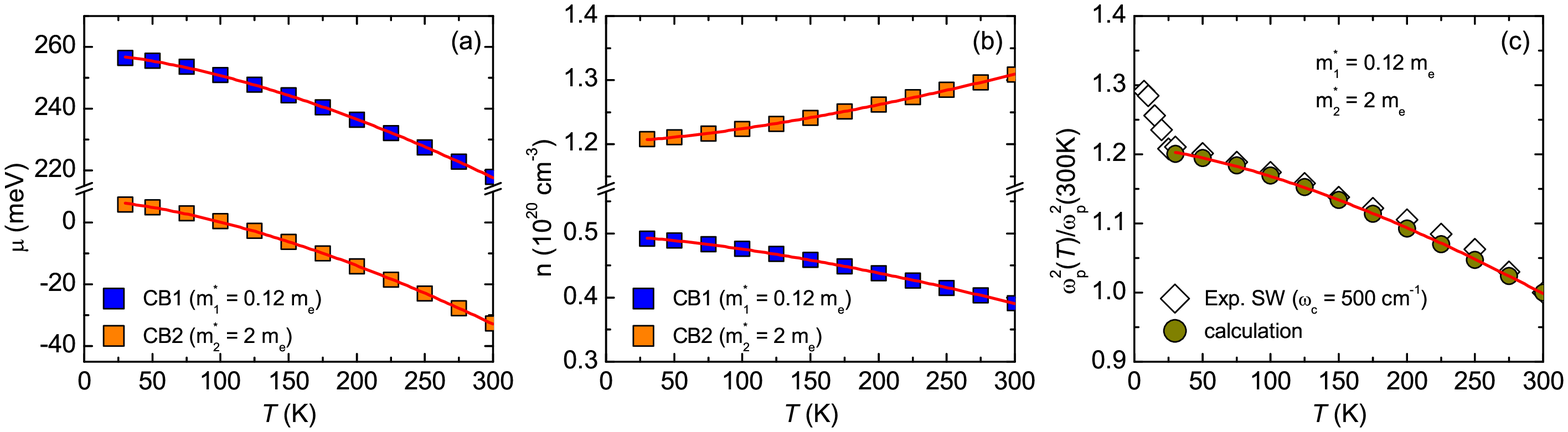}
\caption{ (color online) Effect of a reduced effective mass of the carriers in CB2 of $m = 2~m_e$ on the $T$ dependence of (a) the chemical potentials of CB1 and CB2. (b) the carrier densities of CB1 and CB2, and (c) the ratio $\omega^2_{\mathrm{pl}}(T)/\omega^2_{\mathrm{pl}}(\mathrm{300~K})$.}
\label{FigS11}
\end{figure*}
For the two conduction bands of MnBi$_2$Te$_4$, CB1 and CB2, the total carrier density at finite temperature thus can be written as:
\begin{equation}
\label{total_density}
n(T) = n_1(T) + n_2(T)
\end{equation}
with
\begin{equation}
\label{CB1_density}
n_1(T)= g_sg_b\frac{2}{\sqrt{\pi}}(\frac{m_1^{\ast}kT}{2\pi\hbar^2})^{3/2} \int^{\infty}_{0}\frac{\sqrt{y}}{1 + e^{y-\frac{\mu_1}{kT}}}dy.
\end{equation}
and
\begin{equation}
\label{CB2_density}
n_2(T) = g_sg_b\frac{2}{\sqrt{\pi}}(\frac{m_2^{\ast}kT}{2\pi\hbar^2})^{3/2} \int^{\infty}_{\frac{\mu_1-\mu_2}{kT}}\frac{\sqrt{y}}{1 + e^{y-\frac{\mu_1}{kT}}}dy.
\end{equation}
By substituting $m_1^{\ast} = 0.12~m_e$ and $g_sg_b =4$ into Eq.~(\ref{CB1_density}), as shown in Fig.~\ref{FigS8}(a), we can plot the carrier density of CB1 at 30 and 300 K as a function of $\mu_1$ ($\mu_1 \equiv E_F - E_{CB1}$). In section C, we have obtained the carrier density of CB1 at 30~K of $n_{1}(\mathrm{30~K}) \simeq 0.517 \times 10^{20}~\mathrm{cm}^{-3}$. By using the value $n_1(\mathrm{30~K})$, we can determine $\mu_1(\mathrm{30~K}) = 265$~meV.

Note that in Eq.~(\ref{CB2_density}) $\mu_1 - \mu_2$ defines the relative position of CB2 with respect to CB1. Since in MnBi$_2$Te$_4$ there is a splitting of CB1 below $T_N \simeq 25$~K, in the following calculations we use $\mu_1 (T = \mathrm{30~K}) - \mu_2 (T = \mathrm{30~K})$ as the relative position of CB2 and CB1.

Similarly, by substituting $m_2^{\ast} = 3~m_e$, $g_sg_b =4$, and $\mu_1(\mathrm{30~K}) = 265$~meV into Eq.~(\ref{CB2_density}), we can plot the carrier density of CB2 at 30 K as a function of $\mu_2$, as shown by the blue line in Fig.~\ref{FigS8}(b). Furthermore, by using $n_{2}(\mathrm{30~K}) \simeq 1.183 \times 10^{20}~\mathrm{cm}^{-3}$ obtained in section C, we can determine $\mu_2(\mathrm{30~K}) = 2.37$~meV. As shown in Fig.~\ref{FigS8}(a) and Fig.~\ref{FigS8}(b), it is clear that the thermal broadening of the Fermi-edge has a much stronger influence for the states in CB2 than for the ones in CB1. As sketched in Fig.~\ref{FigS7}, upon increasing the temperature ($T$), the thermal broadening will give rise to a shift of the chemical potential $\Delta \mu (T)$, according to
\begin{equation}
\label{eq_density}
\Delta \mu (T) =  \mu_1 (\mathrm{30~K}) - \mu_1 (T) = \mu_2 (\mathrm{30~K}) - \mu_2 (T)
\end{equation}
and
\begin{equation}
\label{eq_chemicalpotential}
\Delta n (T) = n_1 (\mathrm{30~K}) - n_1 (T) = n_2 (T) - n_2 (\mathrm{30~K}).
\end{equation}
For the measured temperature range between 30 and 300 K, this yields
\begin{widetext}
\begin{equation}
\label{eq_chemicalpotential300K}
\begin{split}
&\frac{n_1 (\mathrm{30~K}) - n_1 (\mathrm{300~K})}{n_2 (\mathrm{30~K}) - n_2 (\mathrm{300~K})} = \\ &\frac{(\frac{m_1^{\ast}kT(\mathrm{30~K})}{2\pi\hbar^2})^{3/2} \int^{\infty}_{0}\frac{\sqrt{y}}{1 + e^{y-\frac{\mu_1(\mathrm{30~K})}{kT(\mathrm{30~K})}}}dy - (\frac{m_1^{\ast}kT(\mathrm{300~K})}{2\pi\hbar^2})^{3/2} \int^{\infty}_{0}\frac{\sqrt{y}}{1 + e^{y-\frac{\mu_1(\mathrm{30~K}) - \Delta \mu(\mathrm{300~K})}{kT(\mathrm{300~K})}}}dy}{(\frac{m_2^{\ast}kT(\mathrm{30~K})}{2\pi\hbar^2})^{3/2} \int^{\infty}_{\frac{\mu_1(\mathrm{30~K})-\mu_2(\mathrm{30~K})}{kT(\mathrm{30~K})}}\frac{\sqrt{y}}{1 + e^{y-\frac{\mu_1(\mathrm{30~K})}{kT(\mathrm{30~K})}}}dy - (\frac{m_2^{\ast}kT(\mathrm{300~K})}{2\pi\hbar^2})^{3/2} \int^{\infty}_{\frac{\mu_1(\mathrm{30~K})-\mu_2(\mathrm{30~K})}{kT(\mathrm{300~K})}}\frac{\sqrt{y}}{1 + e^{y-\frac{\mu_1(\mathrm{30~K}) - \Delta \mu(\mathrm{300~K})}{kT(\mathrm{300~K})}}}dy} = -1.
\end{split}
\end{equation}
\end{widetext}
By substituting $m_1^{\ast} = 0.12~m_e$, $m_2^{\ast} = 3~m_e$, $\mu_1(\mathrm{30~K}) = 265$~meV, and $\mu_2(\mathrm{30~K}) = 2.37$~meV into Eq.~(\ref{eq_chemicalpotential300K}), we obtain $\Delta \mu (\mathrm{300~K}) = 52.37$~meV, and thus $\mu_1 (\mathrm{300~K}) = \mu_1 (\mathrm{30~K}) - \Delta \mu (\mathrm{300~K}) = 212.63$~meV, $\mu_2 (\mathrm{300~K}) = \mu_2 (\mathrm{30~K}) - \Delta \mu (\mathrm{300~K}) = -50$~meV, and $\Delta n (\mathrm{300~K}) = 0.13957 \times 10^{20} \mathrm{cm}^{-3}$ . Figure~\ref{FigS8}(c) and Figure~\ref{FigS8}(d) show the full temperature dependence of the shift of the chemical potential and the subsequent redistribution of carriers between CB1 and CB2.
Accordingly, we obtain the ratio of the corresponding free carrier plasma frequencies of
\begin{equation}
\label{eq_plasma_ratio}
\begin{split}
&\frac{\omega^2_{\mathrm{pl}}(\mathrm{30~K})}{\omega^2_{\mathrm{pl}}(\mathrm{300~K})} = \frac{\omega^2_{\mathrm{pl,1}}(\mathrm{30~K}) + \omega^2_{\mathrm{pl,2}}(\mathrm{30~K})}{\omega^2_{\mathrm{pl,1}}(\mathrm{300~K}) + \omega^2_{\mathrm{pl,2}}(\mathrm{300~K})} = \\
&\frac{\frac{n_1(\mathrm{30~K})}{m_1^{\ast}} + \frac{n_2(\mathrm{30~K})}{m_2^{\ast}}}{\frac{(n_1(\mathrm{30~K}) - \Delta n(\mathrm{300~K}))}{m_1^{\ast}} + \frac{(n_2(\mathrm{30~K}) + \Delta n(\mathrm{300~K}))}{m_2^{\ast}}}.
\end{split}
\end{equation}
By substituting $m_1^{\ast} = 0.12~m_e$, $m_2^{\ast} = 3~m_e$, $n_{1}(\mathrm{30~K}) \simeq 0.517 \times 10^{20}~\mathrm{cm}^{-3}$, $n_{2}(\mathrm{30~K}) \simeq 1.183 \times 10^{20}~\mathrm{cm}^{-3}$ and $\Delta n(\mathrm{300~K}) = 0.13957 \times 10^{20} \mathrm{cm}^{-3}$ into Eq.~(\ref{eq_plasma_ratio}), we thus obtain
\begin{equation}
\frac{\omega^2_{\mathrm{pl}}(\mathrm{30~K})}{\omega^2_{\mathrm{pl}}(\mathrm{300~K})} \simeq 1.3.
\end{equation}
The full temperature dependence of the chemical potentials, the carrier densities and the ratio $\omega^2_{\mathrm{pl}}(T)/\omega^2_{\mathrm{pl}}(\mathrm{300~K})$ is displayed in Fig.~\ref{FigS9}.

These calculations show that the increase of $\omega_{\mathrm{pl}}^2$ toward low $T$, with a ratio between 30 and 300~K of $\frac{\omega^2_{\mathrm{pl}}(\mathrm{30~K})}{\omega^2_{\mathrm{pl}}(\mathrm{300~K})} \simeq 1.20$, can be readily explained in terms of a redistribution of electrons from CB2 to CB1 that arises from sharpening of the Fermi-function and the related increase of the chemical potential by $\Delta \mu = 52.37$~meV.
Note that such a shift of the chemical potential (above $T_N$) is consistent with the $T$ dependence of $E_{dir}$ as shown in Fig.~2(d) in the main text.
The calculated value of the increase of the ratio of the squared plasma frequency is even somewhat larger (by about 10\%) than the one seen in the experiment. The agreement can still be considered as very good, given the simplicity of the two band model used in the calculations which does not take into account the anti-crossing of CB1 and CB2 imposed by the crystal symmetry that is evident in the band calculations~\cite{Zhang2019PRL,Li2019SA} and also shown in the schematics in Fig.~\ref{FigS7}.
It is also possible that the value of the effective mass of the carriers at the bottom of CB2 has been somewhat overestimated. For such a flat band, it can indeed be expected that the parabolic fitting procedure discussed in section E has a sizeable error bar. As shown in Fig.~\ref{FigS10} and Fig.~\ref{FigS11}, we can obtain an excellent  agreement between the experimental and the calculated values of the $T$ dependent increase of the plasma frequency if we reduce the effective mass of CB2 to a value of $m_2^{\ast} = 2~m_e$.
Assuming a value of $m_2^{\ast} = 2~m_e$, instead of $m_2^{\ast} = 3~m_e$, would of course also affect the values of the other calculated parameters. The recalculation, using the relationships $n_{1} + n_{2} = n = 1.7 \times 10^{20}~\mathrm{cm}^{-3}$ and $\omega_{\mathrm{pl}} = \sqrt{\omega^2_{\mathrm{pl},1} + \omega^2_{\mathrm{pl},2}} = \sqrt{n_{1}e^2/\epsilon_0 m^{\ast}_{1} + n_{2}e^2/\epsilon_0 m^{\ast}_{2}} = 6490\icm$ at 30~K and the input parameters $m_1^{\ast} = 0.12~m_e$ and $m_2^{\ast} = 2~m_e$, yields values $n_{1}(\mathrm{30~K}) = 0.492 \times 10^{20}~\mathrm{cm}^{-3}$, $n_{2}(\mathrm{30~K}) = 1.208 \times 10^{20}~\mathrm{cm}^{-3}$, $\mu_1(\mathrm{30~K}) \simeq 256$~meV, $\mu_2(\mathrm{30~K}) \simeq 6$~meV, $\Delta \mu = 38.5$~meV between 30 and 300~K. These are still consistent with our experimental data.

Finally, we remark that at high temperature, although the chemical potential falls slightly below CB2, the states near the bottom of CB2 are still occupied with a high probability due to the thermal broadening of the Fermi-function. It is thus not expected that the interband transitions from the top of VB1 to the bottom of CB2 gives rise to a strong peak in the optical conductivity. The experimental data exhibit indeed only a gradual increase of $\sigma_1(\omega)$ in the relevant frequency range around 3500\icm.

\section{G: Redistribution of electrons between CB1 and CB2 below $T_N$}

With respect to the changes below $T_N$, the thermal effects described in section F yield a 2\% increase of $\omega_{\mathrm{pl}}^2$ that is considerably smaller than the 7\% increase in the experimental data and can also not reproduce the sudden slope change around $T_N$.

As outlined below, the anomalous increase of $\omega_{\mathrm{pl}}^2$ below  $T_N$ can be explained in terms of a small increase of the separation between CB1 and CB2.

To simplify the modeling, we neglect the thermal effects and use here the equations at $T =$ 0~K. With a parabolic band approximation for $\mu = \hbar^2k_F^2/2m^{\ast}$ and the carrier density $n = \frac{1}{(2\pi)^3}\frac{4}{3}\pi k^3_F g_sg_b$, we derive $\mu = \frac{\hbar^2}{2m^\ast}(\frac{6\pi^2 n}{g_sg_b})^{2/3}$ or $n = \frac{g_sg_b}{6 \pi^2}(\frac{2m^{\ast}\mu}{\hbar^2})^{3/2}$. For $m^{\ast}_1 = 0.12~m_e$, $n_{1} \simeq 0.517 \times 10^{20}~\mathrm{cm}^{-3}$, $m^{\ast}_2 = 3~m_e$, and $n_{2} \simeq 1.183 \times 10^{20}~\mathrm{cm}^{-3}$, we obtained $\mu_1 = 266$~meV and $\mu_2 = 19$~meV.

\begin{figure}[t]
\includegraphics[width=0.5\textwidth]{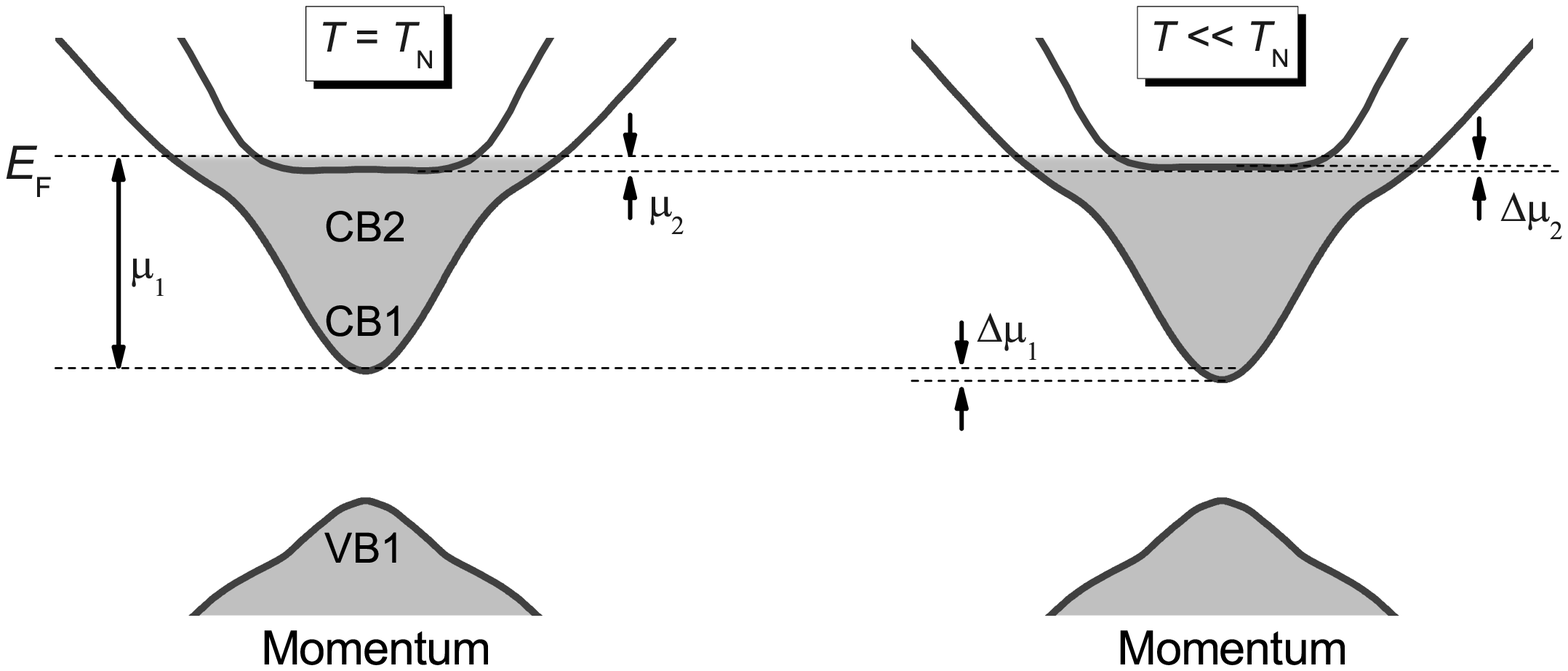}
\caption{ (color online) Schematics of the $T$ dependent shift of the conduction bands CB1 and CB2 below $T_N$. Note that the splitting of CB1  $T_N$ is not shown here.}
\label{FigS12}
\end{figure}
As sketched in Fig.~\ref{FigS12}, for a small increase of the separation between CB1 and CB2 from $T_N$ to $T \ll T_N$, the chemical potentials of CB1 and CB2 are changing from $\mu_1$ and $\mu_2$ at $T_N$ to $\mu_1 + \Delta \mu_1$ and $\mu_2 - \Delta \mu_2$ at $T \ll T_N$. Note that the splitting of CB1 below $T_N$ (into CB1a and CB1b) is not considered here. The resulting transfer of carriers from CB2 to CB1 is estimated according to:
\begin{equation}
\label{eq18}
n_1(\mu_1) - n_1(\mu_1 + \Delta \mu_1) = n_2(\mu_2 - \Delta \mu_2) - n_2(\mu_2).
\end{equation}
To explain an additional 5\% increase of $\omega_{\mathrm{pl}}^2$ below $T_N$ the ratio of the plasma frequencies must be
\begin{equation}
\label{eq19}
\frac{\omega^2_{\mathrm{pl}}(T \ll T_N)}{\omega^2_{\mathrm{pl}}(T \simeq T_N)} = \frac{n_1(\mu_1 + \Delta \mu_1)/m_1^{\ast} + n_2(\mu_2 - \Delta \mu_2)/m_2^{\ast}}{n_1(\mu_1)/m_1^{\ast} + n_2(\mu_2)/m_2^{\ast}} = 1.05.
\end{equation}
By substituting $\mu_1 =$ 266~meV, $m_1^{\ast} = 0.12~m_e$, $\mu_2 =$ 19~meV, and $m_2^{\ast} = 3~m_e$ into equations (\ref{eq18}) and (\ref{eq19}), we thus obtain $\Delta \mu_1 \simeq $ 10~meV and $\Delta \mu_2 \simeq $ 0.3~meV, respectively. Accordingly, the required increase of the separation between CB1 and CB2 below $T_N$ amounts to $\Delta \mu_1 + \Delta \mu_2 \simeq$ 10~meV.

At least, we consider the effect of the splitting of CB1 (into CB1a and CB1b) below  $T_N$ which turns out to be too small and of the wrong sign to explain the experimental trend.

The carrier concentration of CB1 in the paramagnetic state is taken as
\begin{equation}
n_1 =  \frac{g_sg_b}{6 \pi^2}(\frac{2m_1^{\ast}\mu_1}{\hbar^2})^{3/2},
\end{equation}
and in the AFM state (with the exchange splitting into CB1a and CB1b) as
\begin{equation}
n_{1a} + n_{1b} =  \frac{g_sg_b/2}{6 \pi^2}(\frac{2m_{1a}^{\ast}\mu_{1a}}{\hbar^2})^{3/2} + \frac{g_sg_b/2}{6 \pi^2}(\frac{2m_{1b}^{\ast}\mu_{1b}}{\hbar^2})^{3/2}.
\end{equation}
By substituting $\mu_1 =$ 0.266~eV, $m_{1a}^{\ast} = 0.15~m_e$, $m_{1b}^{\ast} = 0.09~m_e$, $\mu_{1a} = 0.266 - 0.027$~eV, $\mu_{1b} = 0.266 + 0.027$~eV, and $g_sg_b =4$, we get $n_{1a} + n_{1b} = 0.501 \times 10^{20}~\mathrm{cm}^{-3} < n_1 = 0.516 \times 10^{20}~\mathrm{cm}^{-3}$.
Assuming a constant carrier concentration of CB1 in the paramagnetic and AFM states, i.e. $n_{1a} + n_{1b} = n_1$, this gives rise to a slight increase of the Fermi-level $\Delta \mu$. In the presence of CB2, which has a much larger effective mass and thus higher density of states at the Fermi-level, such an increase of the Fermi-level will be counteracted by a redistribution of electrons from CB1 to CB2 according to,
$n_{1a}(\mu_{1a} + \Delta \mu) + n_{1b}(\mu_{1b} + \Delta \mu) - n_1(\mu_1) = n_{2}(\mu_2) - n_{2}(\mu_2 + \Delta \mu)$.
By substituting $\mu_1 =$ 0.266~eV, $m_{1a}^{\ast} = 0.15~m_e$, $m_{1b}^{\ast} = 0.09~m_e$, $\mu_{1a} = 0.266 - 0.027$~eV, $\mu_{1b} = 0.266 + 0.027$~eV, $\mu_2 =$ 0.019~eV, and $m_2^{\ast} = 3~m_e$, we get $\Delta \mu = $ 0.15~meV and $\Delta n_1 = - \Delta n_2 = - 0.015 \times 10^{20}~\mathrm{cm}^{-3}$. Since the total plasma frequency is dominated by the much lighter electrons of CB1, the above described effect would give rise to a $\sim$~3\% decrease of $\omega_{\mathrm{pl}}^2$ below $T_N$, as opposed to the observed anomalous increase of $\omega_{\mathrm{pl}}^2$ below $T_N$.

\section{H: Comparison with electron spin resonance (ESR) data on MnBi$_2$Te$_4$}
The Korringa slope of the relaxation rate in the paramagnetic state of electron spin resonance (ESR) measurements is defined as~\cite{Korringa1950,Barnes1981}
\begin{equation}
b=\frac{\pi k_{\mathrm{B}}}{g \mu_{\mathrm{B}}} J_{\mathrm{d}-\mathrm{ce}}^{2} D^{2}(\mu)~(\mathrm{G}/\mathrm{K})
\end{equation}
where $k_B = 8.62 \times 10^{-5}$~eV/K, $\mu_B = 5.79 \times 10^{-9}$~eV/G and $g = 2$ (for the Mn$^{2+}$ ion), $J_{\mathrm{d}-\mathrm{ce}}$ is the exchange coupling constant between the spins of the localized Mn $3d$ and the conduction electrons, and $D(\mu)$ is the density of states of the conduction electrons at the Fermi level. The magnitude of $J_{\mathrm{d}-\mathrm{ce}}$ thus can be estimated using the values of $b$ and $D(\mu)$ as obtained from the ESR and optical data, \begin{equation}
J_{\mathrm{d}-\mathrm{ce}}=D^{-1}(\mu) \sqrt{\frac{b}{\pi k_{\mathrm{B}}/g \mu_{\mathrm{B}}}}.
\end{equation}
$D(\mu)$ can be deduced according to
\begin{equation}
D(\mu)= \frac{V}{4\pi^2} (\frac{2m^{\ast}}{\hbar^2})^{3/2} \sqrt{\mu}~g_sg_b,
\end{equation} using the values of the effective mass, $m^{\ast}$ and $\mu$ as obtained from the optical data and the unit cell volume $V = 665.1~\mathrm{\AA}^{3} = 6.651 \times 10^{-28}~\mathrm{m}^{3}$ taken from Ref.~\cite{Zeugner2019CM}. In addition, we have to account for the difference in the carrier concentrations (as determined from the Hall-effect) of the crystals on which the ESR and optics experiments have been performed.
For the crystal A of the optics experiment we obtained $m_1^{\ast} \approx 0.12~m_e$, $m_2^{\ast} \approx 3~m_e$ and $\mu_1 \approx$ 0.266~eV, $\mu_2 \approx$ 0.019~eV. Considering the spin- and band degeneracy $g_s = 2$ and $g_b =2$, respectively, of CB1 and CB2 we thus obtain $D(\mu_1 = 0.266~\mathrm{eV}) =$ 0.19 states/eV/fu (formula unit) and $D(\mu_2 = 0.019~\mathrm{eV}) =$ 6.47 states/eV/fu for CB1 and CB2, respectively.
Taking into account the lower carrier concentration of $n = 2 \times 10^{19}$~cm$^{-3}$~\cite{Otrokov2019Nat}, as compared to $n = 1.7 \times 10^{20}$~cm$^{-3}$ of our optics crystal A, and using the relationship $\mu = (\hbar^2/2m^\ast)(6\pi^2 n/g_sg_b)^{2/3}$, we estimate that the Fermi-level of the ESR sample crosses CB1 at $\mu = 0.15$~eV, whereas CB2 remains above the Fermi-level (and thus empty). Note that this estimate agrees with the ARPES data of the ESR sample that are shown in Ref.~\cite{Otrokov2019Nat}.
With $m^{\ast} = m_1^{\ast} \approx 0.12~m_e$, we thus obtain $D(\mu= 0.15~eV) =$ 0.15 states/eV/fu.
The ESR experiment reported in Ref.~\cite{Otrokov2019Nat}, revealed a Korringa-relation with a linear slope of the relaxation rate of $b^\perp = 2.4$ G/K for $H \perp c$ axis and $b^\parallel = 7.3$ G/K for $H \parallel c$ axis, respectively. Accordingly, we derive values of $J_{\mathrm{d}-\mathrm{ce}} \simeq$ 70 and 120~meV, respectively, that are of the same order of magnitude as the band splitting between CB1a and CB1b in our optical data of 54 meV. This reasonable agreement, given the uncertainties due to the rather large difference in the carrier concentrations of the samples studied with ESR and optics and the experimental error bars of the various measured parameters, confirms our interpretation of the optical data in terms of an exchange splitting of CB1 that is much larger than the one of CB2 (which has been neglected in the main text). The much weaker exchange coupling of the electrons in CB2 with the Mn spins can also explain that despite the rather different concentration of conduction electrons the various MBT samples have very similar values of $T_N$~\cite{Yan2019PRM,Cui2019PRB,Lee2019PRR,Otrokov2019Nat,Chen2019PRM,Li2020PCCP}, since the former changes in carrier concentration mainly affect the occupation of CB2.

%
%

\end{document}